\documentclass[preprints,review,accept,pdftex,moreauthors]{Definitions/mdpi} 

\firstpage{1} 
\pubvolume{1}
\issuenum{1}
\articlenumber{0}
\pubyear{2023}
\copyrightyear{2023}
\datereceived{ } 
\daterevised{ } 
\dateaccepted{ } 
\datepublished{ } 
\hreflink{https://doi.org/} 

\usepackage{subfigure}
\usepackage{xspace}
\usepackage{epsfig}
\usepackage{rotating}
\usepackage{dcolumn}
\usepackage{overpic}

\newboolean{pdflatex}
\setboolean{pdflatex}{true} 

\newboolean{articletitles}
\setboolean{articletitles}{true} 

\newboolean{uprightparticles}
\setboolean{uprightparticles}{false} 
\usepackage{amsfonts}
\usepackage{bm}
\usepackage{ctable}
\usepackage{placeins}
\usepackage{comment}

\def\st     {ST\xspace}

\ifthenelse{\boolean{uprightparticles}}%
{

 \def\Pmu         {\ensuremath{\upmu}\xspace}                 
 \def\Pnu         {\ensuremath{\upnu}\xspace}                 
                  
 \def\Ppi         {\ensuremath{\uppi}\xspace}

 \def\Ppsi        {\ensuremath{\uppsi}\xspace}

 \def\PDelta      {\ensuremath{\Delta}\xspace}                 
 \def\PXi      {\ensuremath{\Xi}\xspace}                 
 \def\PLambda      {\ensuremath{\Lambda}\xspace}                 
 \def\PSigma      {\ensuremath{\Sigma}\xspace}                 
 \def\POmega      {\ensuremath{\Omega}\xspace}                 
 \def\PUpsilon      {\ensuremath{\Upsilon}\xspace}                 
 
 \def\PB      {\ensuremath{\mathrm{B}}\xspace}                 
                  
 \def\PD      {\ensuremath{\mathrm{D}}\xspace}

 \def\PJ      {\ensuremath{\mathrm{J}}\xspace}                 
 \def\PK      {\ensuremath{\mathrm{K}}\xspace}

 \def\Pb      {\ensuremath{\mathrm{b}}\xspace}                 
 \def\Pc      {\ensuremath{\mathrm{c}}\xspace}

 \def\Ph      {\ensuremath{\mathrm{h}}\xspace}                 
 \def\Pi      {\ensuremath{\mathrm{i}}\xspace}

 \def\Pp      {\ensuremath{\mathrm{p}}\xspace}

 \def\Ps      {\ensuremath{\mathrm{s}}\xspace}

}
{

 \def\Pmu         {\ensuremath{\mu}\xspace}                 
 \def\Pnu         {\ensuremath{\nu}\xspace}                 
                  
 \def\Ppi         {\ensuremath{\pi}\xspace}

 \def\Ppsi        {\ensuremath{\psi}\xspace}                 
                  
 \mathchardef\PDelta="7101
 \mathchardef\PXi="7104
 \mathchardef\PLambda="7103
 \mathchardef\PSigma="7106
 \mathchardef\POmega="710A
 \mathchardef\PUpsilon="7107
                  
 \def\PB      {\ensuremath{B}\xspace}                 
                  
 \def\PD      {\ensuremath{D}\xspace}

 \def\PJ      {\ensuremath{J}\xspace}                 
 \def\PK      {\ensuremath{K}\xspace}

 \def\Pb      {\ensuremath{b}\xspace}                 
 \def\Pc      {\ensuremath{c}\xspace}

 \def\Ph      {\ensuremath{h}\xspace}                 
 \def\Pi      {\ensuremath{i}\xspace}

 \def\Pp      {\ensuremath{p}\xspace}

 \def\Ps      {\ensuremath{s}\xspace}

}

\def\muon       {\ensuremath{\Pmu}\xspace}
\def\mup        {\ensuremath{\Pmu^+}\xspace}
\def\mun        {\ensuremath{\Pmu^-}\xspace} 
\def\mumu       {\ensuremath{\Pmu^+\Pmu^-}\xspace}

\def\neu        {\ensuremath{\Pnu}\xspace}

\def\neum       {\ensuremath{\neu_\mu}\xspace}

\def\squark    {\ensuremath{\Ps}\xspace}

\def\cquark    {\ensuremath{\Pc}\xspace}

\def\bquark    {\ensuremath{\Pb}\xspace}

\def\hadron {\ensuremath{\Ph}\xspace}
\def\pion  {\ensuremath{\Ppi}\xspace}

\def\pip   {\ensuremath{\pion^+}\xspace}
\def\pim   {\ensuremath{\pion^-}\xspace}

\def\kaon  {\ensuremath{\PK}\xspace}
  \def\Kbar  {\kern 0.2em\overline{\kern -0.2em \PK}{}\xspace}

\def\Kp    {\ensuremath{\kaon^+}\xspace}
\def\Km    {\ensuremath{\kaon^-}\xspace}

  \def\Dbar    {\kern 0.2em\overline{\kern -0.2em \PD}{}\xspace}
\def\D       {\ensuremath{\PD}\xspace}

\def\Dz      {\ensuremath{\D^0}\xspace}

\def\Dstarp  {\ensuremath{\D^{*+}}\xspace}

\def\B       {\ensuremath{\PB}\xspace}
\def\Bbar    {\ensuremath{\kern 0.18em\overline{\kern -0.18em \PB}{}}\xspace}

\def\Bz      {\ensuremath{\B^0}\xspace}

\def\Bu      {\ensuremath{\B^+}\xspace}

\def\Bd      {\ensuremath{\B^0}\xspace}
\def\Bs      {\ensuremath{\B^0_\squark}\xspace}
\def\Bsb     {\ensuremath{\Bbar^0_\squark}\xspace}

\def\Bc      {\ensuremath{\B_\cquark^+}\xspace}
\def\Bcp     {\ensuremath{\B_\cquark^+}\xspace}

\def\jpsi     {\ensuremath{{\PJ\mskip -3mu/\mskip -2mu\Ppsi\mskip 2mu}}\xspace}

  \def\Y#1S{\ensuremath{\PUpsilon{(#1S)}}\xspace}

\def\proton      {\ensuremath{\Pp}\xspace}

\def\Lz {\ensuremath{\PLambda}\xspace}
\def\Lbar {\ensuremath{\kern 0.1em\overline{\kern -0.1em\PLambda}}\xspace}

\def\Lb      {\ensuremath{\Lz^0_\bquark}\xspace}

\def\BR         {\BF}
\newcommand{\decay}[2]{\ensuremath{#1\!\to #2}\xspace}         

\def\to                 {\ensuremath{\rightarrow}\xspace}

\def\CP                {\ensuremath{C\!P}\xspace}

\def\BsToJPsiPhi  {\decay{\Bs}{\jpsi\phi}}

\def\AT#1     {\ensuremath{A_{\mathrm{T}}^{#1}}\xspace}           

\def\C#1      {\ensuremath{\mathcal{C}_{#1}}\xspace}                       
\def\Cp#1     {\ensuremath{\mathcal{C}_{#1}^{'}}\xspace}                    
\def\Ceff#1   {\ensuremath{\mathcal{C}_{#1}^{\mathrm{(eff)}}}\xspace}        
\def\Cpeff#1  {\ensuremath{\mathcal{C}_{#1}^{'\mathrm{(eff)}}}\xspace}       
\def\Ope#1    {\ensuremath{\mathcal{O}_{#1}}\xspace}                       
\def\Opep#1   {\ensuremath{\mathcal{O}_{#1}^{'}}\xspace}                    

\newcommand{\tev}{\ensuremath{\mathrm{\,Te\kern -0.1em V}}\xspace}
\newcommand{\gev}{\ensuremath{\mathrm{\,Ge\kern -0.1em V}}\xspace}
\newcommand{\mev}{\ensuremath{\mathrm{\,Me\kern -0.1em V}}\xspace}
\newcommand{\kev}{\ensuremath{\mathrm{\,ke\kern -0.1em V}}\xspace}
\newcommand{\ev}{\ensuremath{\mathrm{\,e\kern -0.1em V}}\xspace}
\newcommand{\gevc}{\ensuremath{{\mathrm{\,Ge\kern -0.1em V\!/}c}}\xspace}
\newcommand{\mevc}{\ensuremath{{\mathrm{\,Me\kern -0.1em V\!/}c}}\xspace}
\newcommand{\gevcc}{\ensuremath{{\mathrm{\,Ge\kern -0.1em V\!/}c^2}}\xspace}
\newcommand{\gevgevcccc}{\ensuremath{{\mathrm{\,Ge\kern -0.1em V^2\!/}c^4}}\xspace}
\newcommand{\mevcc}{\ensuremath{{\mathrm{\,Me\kern -0.1em V\!/}c^2}}\xspace}

\def\invfb   {\ensuremath{\mbox{\,fb}^{-1}}\xspace}

\def\ps   {\ensuremath{{\rm \,ps}}\xspace}

\def\gsim{{~\raise.15em\hbox{$>$}\kern-.85em
          \lower.35em\hbox{$\sim$}~}\xspace}
\def\lsim{{~\raise.15em\hbox{$<$}\kern-.85em
          \lower.35em\hbox{$\sim$}~}\xspace}

\def\sPlot{\mbox{\em sPlot}}

\def\sqs   {\ensuremath{\protect\sqrt{s}}\xspace}

\def\pt         {\mbox{$p_{\rm T}$}\xspace}

\def\tell1  {TELL1\xspace}
\def\ukl1   {UKL1\xspace}

\def\BuToJPsiK {\decay{\Bu}{\jpsi\Kp}\xspace}
\def\bujpsik{\BuToJPsiK}

\newcommand{\BF}{branching fraction\xspace}

\newcommand{\taumumu}{\ensuremath{\tau_{\mu\mu}}}
\newcommand{\taumumuSM}{\ensuremath{\tau^{SM}_{\mu\mu}}}
\newcommand{\taumumuobs}{\ensuremath{\tau_{\mu\mu}^\mathrm{Obs}}}

\newcommand{\Bsd}{\ensuremath{B^0_{(s)}}\xspace}

\newcommand{\Bmumu}{\ensuremath{\Bd_{(s)}\to\mu^+\mu^-}\xspace}
\newcommand{\Bsmumu}{\ensuremath{\Bs\to\mu^+\mu^-}\xspace}
\newcommand{\Bdmumu}{\ensuremath{\Bd\to\mu^+\mu^-}\xspace}
\newcommand{\Bsdmumu}{\ensuremath{B_{(s)}^0\to\mu^+\mu^-}\xspace}
\def\bsmumu{\Bsmumu}
\def\bdmumu{\Bdmumu}
\def\bsdmumu{\Bsdmumu}

\newcommand{\Bsmumugamma}{\ensuremath{\Bs\to\mu^+\mu^-\gamma}\xspace}

\newcommand{\BsKK}{\ensuremath{\Bs\to K^+K^-}\xspace}

\newcommand{\BdKpi}{\ensuremath{\Bd\to K^+\pi^-}\xspace}

\newcommand{\Bhh}{\ensuremath{B^0_{(s)}\to h^+h^-}\xspace}

\newcommand{\Bhhprime}{\ensuremath{B^0_{(s)}\to h^+h^{(')-}}\xspace}

\newcommand{\BpmJpsiK}{\ensuremath{B^{\pm}\to J/\psi K^{\pm}}\xspace}
\newcommand{\BuJpsimumuK}{\ensuremath{B^+\to J/\psi(\mu^+\mu^-)K^+}\xspace}

\newcommand{\ys}{\ensuremath{y_s}\xspace}
\newcommand{\BAR}[1]{\overline{#1}}
\newcommand{\Bsbar}{\ensuremath{\BAR{B}^0_s}\xspace}

\newcommand{\BRof}[1]{\ensuremath{{\cal B}(#1)}\xspace}
\def\ADeltaGamma{\ensuremath{\mathcal{A}_{\Delta\Gamma}}\xspace}
\def\ys{\ensuremath{y_s}}
\def\DeltaGammas{\ensuremath{\Delta\Gamma_s}}

\def\RB{\ensuremath{{\mathcal{R}}}\xspace}

\newcommand{\BRBsmm}{\ensuremath{3.66\pm0.14}\xspace}
\newcommand{\BRBdmm}{\ensuremath{1.03\pm0.05}\xspace}
\newcommand{\RBSM}{\ensuremath{0.0281 \pm 0.0016}\xspace} 

\newcommand{\eN}[1]{\ensuremath{#1 \times 10^{-9}}\xspace}
\newcommand{\eT}[1]{\ensuremath{#1 \times 10^{-10}}\xspace}

\def\LHCbBsmumuMeasuresyst{\ensuremath{\left(3.0 \pm 0.6^{\,+0.3}_{\,-0.2}\right)\times 10^{-9}}\xspace}
\def\LHCbBsmumuSignificance{\ensuremath{7.8}\xspace}
\def\LHCbBdmumuMeasure{\ensuremath{\left(1.5^{\,+1.2\,+0.2}_{\,-1.0\,-0.1}\right)\times 10^{-10}}\xspace}
\def\LHCbBdmumuSignificance{\ensuremath{1.6}\xspace}
\def\LHCbBdmumuUpperLimit{\ensuremath{3.4\times 10^{-10}}\xspace} 
\def\LHCbBsmumuEffectiveLifetime{\ensuremath{2.04\pm 0.44 \pm 0.05 \,\rm{ps}}\xspace}

\def\CMSBsmumuEffectiveLifetime{\ensuremath{1.70^{+0.61}_{-0.44} \,\rm{ps}}\xspace}

\Title{Analysis of $B^0_{(s)}\to\mu^+\mu^-$ decays at the LHC}

\TitleCitation{Title}

\Author{Kai-Feng Chen $^{1,\dagger}$*\orcidA{}, Titus Momb\"acher $^{2,\dagger}$*\orcidB{} and Umberto de Sanctis $^{3,\dagger}$*\orcidC{}}

\AuthorNames{Kai-Feng Chen, Titus Momb\"acher and Umberto de Sanctis}

\AuthorCitation{Chen, K.-F.; Momb\"acher, T.; de Sanctis, U.}

\address{%
$^{1}$ \quad National Taiwan University\\
$^{2}$ \quad CERN\\
$^{3}$ \quad Universit\`a degli Studi di Roma - Tor Vergata and INFN Sezione di Roma - Tor Vergata}

\corres{Correspondence: KC: Kai-Feng.Chen@cern.ch; TM: titus.mombacher@cern.ch; US: umberto.de.sanctis@uniroma2.it}

\firstnote{The authors contributed equally to this work.}

\abstract{This article reviews the most recent measurements of \Bsdmumu decay properties at the LHC, which are the most precise to date. The measurements of the branching fraction and effective lifetime of the \Bsmumu decay by the ATLAS, CMS and LHCb collaborations, as well as the search for \Bdmumu decays are summarised with a focus on the experimental challenges. Furthermore prospects are given for these measurements and new observables that become accessible with the foreseen amounts of data by the end of the LHC.}

\keyword{\Bsdmumu, FCNC, Rare Decay, Flavour physics, ATLAS, CMS, LHCb, LHC} 

\begin{document}


\section{Introduction}
\label{sec:introduction}
This review summarises the most recent measurements related to the \Bdmumu\ and \Bsmumu\ decays performed with the ATLAS, CMS and LHCb experiments. 

The \Bdmumu\ and \Bsmumu\ decays belong to the category of Flavour Changing Neutral Current
(FCNC) processes and therefore highly suppressed in the Standard Model (SM). This makes them important tools in the search for New Physics (NP) since they can provide indirect constraints on NP processes that interfere with the SM processes and alter the rates and decay properties sizeably. They are even sensitive to particles that are out of the kinematic range accessible by particle colliders, including the Large Hadron Collider (LHC).
The \Bsdmumu decays are among the most sensitive FCNC processes due to their small theoretical uncertainty and clean experimental signature~\cite{Bobeth:2013uxa,Bobeth:2013tba,Hermann:2013kca,Beneke:2017vpq,Beneke:2019slt}. In the SM, the decays \Bsdmumu are forbidden at leading-order and can proceed only
via loop diagrams. In addition, they are also suppressed by the helicity conservation and the presence of off-diagonal CKM matrix elements, leading to very small expected decay time integrated branching fractions.
Additional interest in \Bsmumu decays comes from the simple description in effective field theory~\cite{Altmannshofer:2011gn,Beaujean:2012uj}.
The decays can only proceed via axial-vector (Wilson coefficient $\mathcal{C}_{10}$), scalar ($\mathcal{C}_S$) or pseudo-scalar ($\mathcal{C}_P$) $\bquark\to\squark\ell\bar{\ell}$ currents, where the scalar and pseudo-scalar currents are forbidden in the SM. Thus, measurements of \Bsmumu properties are crucial inputs for global fits of the parameters that govern $\bquark\to\squark\ell\bar{\ell}$ transitions.

The most up-to-date SM predictions for the \Bsmumu and \Bdmumu branching fractions 
are calculated in Ref.~\cite{Beneke:2019slt} and yield
\begin{equation}
\label{eq:sm_prediction}
\begin{array}{lll}
\BRof{\Bsmumu} &=& \eN{(\BRBsmm)} \, \quad \textrm{and}\\
\BRof{\Bdmumu} &=& \eT{(\BRBdmm)}\, \quad .
\end{array}
\end{equation}
The values of these branching fractions are \CP\-averaged and time-integrated and include final state radiation effects, so that they can be readily compared with the experimental measurements, which do not distinguish between $B^0_{(s)}$ \CP\-eigenstates and take final state radiation effects into account.
Next-to-leading order electroweak corrections and  next-to-next-to-leading order QCD corrections are also included in the calculations. 
Recently, several progresses in lattice quantum chromodynamics (QCD)~\cite{Aoki:2019cca,Bazavov:2017lyh,Bussone:2016iua,Dowdall:2013tga,Hughes:2017spc}, 
in the calculation of electroweak effects at next-to-leading order~\cite{Bobeth:2013tba}, 
and QCD effects at next-to-next-to-leading order~\cite{Hermann:2013kca} helped significantly in reducing the theoretical uncertainties on both branching fractions. 
Enhanced electromagnetic contributions from virtual photon exchange were also proven to produce larger corrections to theoretical uncertainties than previously assumed~\cite{Beneke:2017vpq,Beneke:2019slt}.
These predictions also take into account the finite width difference measured in the \Bs system, that apply in the experimental measurements where the data samples are un-tagged (see Ref.~\cite{DeBruyn:2012wk,DeBruyn:2012wj}).
Alternative predictions are also available. They are obtained using the relation between \bsdmumu decays and  $\Delta m_{d(s)}$, the mass difference of the $B^0_{(s)}$ mass eigenstates~\cite{Buras:2003td, King:2019lal}. In addition, it has been recently pointed out~\cite{Buras:2022} that the current way to calculate \BRof{\bsdmumu} could be affected by the presence on NP effects. Therefore, a calculation based on $\Delta m_{d(s)}$ and $|\epsilon_K|$ considering only the SM contribution has been proposed. 
In both cases, the resulting values for \BRof{\bsdmumu} are slightly different than the ones shown in Eq.~\ref{eq:sm_prediction}, but still compatible within the theoretical uncertainties of the calculation, that are, on their own, still smaller than the experimental precision. Since all Collaborations used the values reported in Eq.~\ref{eq:sm_prediction} to assess the level of compatibility of the measurements with the SM predictions, in the remainder of the article, the values quoted in Eq.~\ref{eq:sm_prediction} will be used as reference values for the \BRof{\bsdmumu} SM predictions.

While the mentioned reference does not quote a value for the ratio of the two branching fractions, this can be easily calculated as:
\begin{equation}
\label{eq:RB}
\RB = \frac{\BRof{\Bdmumu}}{\BRof{\Bsmumu}}   
= \frac{\tau_{B^0}}{1/\Gamma_H^{s}} \left(\frac{f_{B^0}}{f_{B^0_s}}\right)^2 
 \left|\frac{V_{td}}{V_{ts}}\right|^2 \tfrac{M_{B^0} \sqrt{1 - \tfrac{4 
 m_{\mu}^2}{M_{B^0}^2}}}{M_{B^0_s} \sqrt{1 - \tfrac{4 m_{\mu}^2}{M_{B^0_s}^2}}}
= \RBSM\quad 
\end{equation}
where $\tau_{B^0}$ and $1/\Gamma_H^{s}$ are the lifetimes of the \Bd and of the heavy mass eigenstate 
of the \Bs; $M_{\Bs}$ and $M_{\Bd}$ are the masses and $f_{\Bs}$ and $f_{\Bd}$ the meson decay constants 
of the \Bs and \Bd mesons respectively; $V_{td}$ and $V_{ts}$ the elements of the CKM matrix and $m_{\mu}$ the mass of the muon. 
Using the same input values as Ref.\cite{Beneke:2019slt}, the numerical value in Eq.~\ref{eq:RB} is obtained. 
It is worth noting that the ratio has a theoretical uncertainty which is smaller than the 
single branching fractions due to the cancellation of most of the factors. 
In particular this ratio has the same value in all theories obeying the Minimal Flavour Violation (MFV) paradigm (including the SM) and as 
such it is a test of the latter. It is therefore of additional interest to evaluate \RB also in upcoming measurements.\\


A second observable of the \bsmumu decay considered in the latest experimental results is its effective lifetime \taumumu . This observable is complementary to the branching fraction because it is sensitive to potential New Physics (NP) effects which are flavour-dependent~\cite{DeBruyn:2012wk}. In fact, in the SM, only the heavy \CP\-odd heavy-mass eigenstate component of the \Bs-\Bsbar system contributes to the \Bsmumu\ decay amplitude: an assumption which does not hold in every NP scenario. Therefore, the measurement of this quantity could reveal the presence of NP effects which do not affect the branching fraction measurement. 
\taumumu\ is simply defined as the mean lifetime of \Bsmumu decays 
\begin{eqnarray}
 \taumumu &\equiv& \frac{\int^\infty_0 t \left<\Gamma \left( \bsmumu \right) \right> dt }{\int^\infty_0 \left<\Gamma \left( \bsmumu \right) \right> dt } \nonumber\\
 &=& \frac{\tau_{B^0_s}}{1 - \ys^2} \left[ \frac{1+ 2 \ADeltaGamma \ys + \ys^2}{1 + \ADeltaGamma \ys} \right],
\label{eq:lifetime}\end{eqnarray}
where $t$ is the proper decay time of the \Bs meson and \ys\ and the \CP parameter \ADeltaGamma are defined as
\begin{equation}
  \ys \equiv \frac{\DeltaGammas}{2 \Gamma_s} \qquad 
  \ADeltaGamma \equiv \frac{R_H^{\mu^+ \mu^-} - R_L^{\mu^+ \mu^-}}{R_H^{\mu^+ \mu^-} + R_L^{\mu^+ \mu^-}} 	
\end{equation}
and $R_H^{\mup \mun}$ and $R_L^{\mup \mun}$ denote the contributions of the heavy and light mass eigenstates of the \Bs system to the un-tagged \Bsmumu decay rate.
Since the \mumu final state is \CP-odd, in the SM $\ADeltaGamma = +1$ and the effective lifetime is equal to the lifetime of the heavy-mass \Bs eigenstate \taumumuSM. The \CP asymmetry \ADeltaGamma can receive contributions from NP effects, particularly from scalar and pseudoscalar operators, even in cases where the branching fraction \BRof{\Bsmumu} is not modified. The most recent \taumumuSM\ value is $ 1.624 \pm 0.009 \ps$~\cite{HFLAV:2022pwe}, which can be slightly different from that used by the various Collaborations depending on publication time of their most recent measurement. 

All experimental results described in this review assume the SM hypothesis $\ADeltaGamma = 1$ in the calculation of efficiencies and acceptance for the \bsmumu\ decay and thus for its branching fractions. However, the ATLAS, CMS and LHCb Collaborations estimated the impact of such assumption on the \bsmumu\ branching fraction, which spans from 4\% to 10\% depending on $\ADeltaGamma$ varying in the interval [-1,1].

The article is structured as follows: in Sections~\ref{sec:atlas_ana},  \ref{sec:cms_BF} and \ref{sec:LHCb_measurements}, the measurements by the ATLAS, CMS and LHCb Collaborations are reported respectively, while in Section \ref{sec:combination} the results obtained by the latest official LHC combination are presented. Section~\ref{sec:prospects} provides a summary of the status of the measurements and prospects of the three Collaborations for the HL-LHC phase.

\section{The ATLAS \boldmath \BRof{\Bsdmumu} and effective lifetime measurements}
\label{sec:atlas_ana}
\subsection{The \BRof{\Bsdmumu} measurement}
The measurement of the branching fractions of the \Bsmumu and \Bdmumu
decays performed by the ATLAS Collaboration is described and
documented in Ref.~\cite{Aaboud:2018mst}.
The analysis uses 26.1~\invfb of Run-2 data collected at \sqs~=~13~\tev,
and combines the result with the previously published Run-1 analysis~\cite{Aaboud:2016ire}
on 4.7~\invfb of data at \sqs~=~7~\tev and 20.3~\invfb at \sqs~=~8~\tev.

In order to remove the dependence from the knowledge of the $b$-quark production
cross section and minimise the systematic uncertainties, the branching fractions 
are measured relative to a reference channel. For its abundance and  well-measured branching ratio, the \BuJpsimumuK decay channel has been chosen for this purpose. 
As a consequence, the procedure to extract the \BRof{\Bsdmumu} takes into account the difference in the fragmentation fractions $f_{u,d,s}$ of \bquark-quarks to form, respectively, 
a \Bu, \Bs or \Bd meson. Also the different acceptances
and efficiencies between the signal and the reference channels are taken into account. 
Hence, the branching fractions \BRof{\Bsdmumu} are expressed as:
\begin{equation}
\BRof{\Bsdmumu} =  {N_{\Bsd} \over N_{B^+}} {f_u \over 
f_{d(s)}} {\epsilon_{\rm tot}^{B^+} \over \epsilon_{\rm tot}} \BRof{\BuJpsimumuK}~,
\label{eq:ATLASMasterFormula}
\end{equation}
where $N_{\Bsd}$ ($N_{B^+}$) is the measured yield of \Bsdmumu ($B^+\to J/\psi K^+$) events,
and $\epsilon_{\rm tot}$  ($\epsilon_{\rm tot}^{B^+}$) is the total signal ($B^+\to J/\psi K^+$) efficiency. 
Events from \BsToJPsiPhi decay, with $J/\psi \to \mu \mu$ and $\phi \to KK$, are also used
as control sample for the signal kinematic variables exploited in the analysis.

The signal selection starts with a hardware dimuon trigger requiring
one muon with transverse momentum \pt~$>$~4~\gev and the other with \pt~$>$~6~\gev.
In the offline analysis, both muons are required to have the same \pt thresholds as in the trigger selection, to have pseudo-rapidity $|\eta|~<~$2.5, and to pass stringent track-quality requirements (\textit{Tight} muons).
Signal candidates are formed with two muons with opposite electric charges.
Kaon candidates for the reference channel are reconstructed in the tracking system and are required to have \pt~$>$~1~\gev and $|\eta|~<~$2.5. 

$B$-meson kinematic observables are reconstructed imposing quality requirements on 
the dimuon vertex for the signal, or on the vertex formed by the dimuon system and
one track for the reference channel. 
The reconstructed $B$ candidates are also required to fall within a fiducial volume
defined as $\pt(B)$~$>$~8~\gev and $|\eta(B)|~<~$2.5. 

The analysis uses mainly the $B$-candidate invariant mass to characterise the selected events. $B$-candidates with a mass in the 4766-5966 MeV interval, are considered.
A blind analysis is performed where the dimuon invariant mass signal region between 
5166 and 5526~\mev is not used until the analysis criteria and strategies are finalised.


The main backgrounds for this analysis can be split into three categories:
continuum background, partially reconstructed $B$ decays (PRD) and peaking background.
The continuum background consists mainly of muons produced in uncorrelated hadron decays.   
It is the dominant background for the analysis and it is several orders of magnitude larger than the signal. Therefore, a Boosted Decision Tree~\cite{Hocker:2007ht} (c-BDT)
is emplyed to efficiently separate the signal from this background type. 
The c-BDT is based on 15 kinematic variables with high discriminating power which describe the kinematics of the $B$-meson candidate, the secondary vertex displacement, 
the kinematic properties of the muons and the rest of the event (such as the isolation of the $B$ candidate and that of the two muon tracks with respect to the rest of the event).
The c-BDT is trained and validated on the data mass sidebands. 

The PRD background is made of decays where the two muons in the final state come from one of the following topologies: $(a)$ 'cascade' transitions with the muons coming from the same ancestor
(e.g. $b \to c \mu \nu \to s \mu \mu \nu \nu$), and labelled {\it same-side} muons (SS); 
$(b)$ from the same decay chain (e.g. $B \to J/\psi X$ or $\Bd \to \mu \mu K^*$) and labelled {\it same vertex} muons (SV); 
$(c)$ from $B_c \to J/\psi \mu \nu$ 
decays; 
$(d)$ from semileptonic $B$ decays where a hadron $h$ ($\pi$, $K$ or proton) is misidentified as a muon (e.g. $B \to \mu h \nu$). 
All these types of backgrounds populate the low-mass sideband with contributions also into the dimuon mass signal region.

The peaking background consists of charmless two body decays \Bhhprime 
 ($h^{(\prime)}$ being a pion or a kaon) that are reconstructed as signal events due to the hadrons being misidentified as muons.
This background component falls in the signal region and presents the same features of the \Bdmumu signal.
Its contribution has been studied with the help of a dedicated MC sample and validated in data in a region enriched by hadrons misidentified as muons.  
The resulting peaking background contribution is estimated to be $2.9 \pm 2.0$ events in the signal region. 

To extract the \BRof{\Bsdmumu} using Eq.~\ref{eq:ATLASMasterFormula}, the 
yield of the reference channel and the efficiency ratio between the two channels needs to be computed.
The \BuToJPsiK yield $N_{B^+}$ is obtained by an unbinned extended maximum-likelihood fit to the $\mu \mu K^+$ invariant mass distribution, where the shape parameters are fitted simultaneously in data and simulation. 

The efficiency ratio between signal and reference channels is computed from appropriate simulation samples within the fiducial volume of the analysis.
These samples are reweighted in such a way that they reproduce the distributions of the number of primary vertices (and therefore pile-up), $\pt(B)$, $|\eta(B)|$ and trigger efficiencies (as a function of $\pt(\mu)$ and $|\eta(\mu)|$) as measured in data.
Furthermore, a correction to the $\Bs$ lifetime in the simulated signal 
sample is applied to match the distribution of the heavy \Bs mass eigenstate, because the \Bsmumu decay proceeds in the SM exclusively through the heavy $\Bs$ mass eigenstate, as described in Section \ref{sec:introduction}. 

The yields of signal events $N_{\Bsd}$ are extracted simultaneously from an unbinned extended maximum-likelihood fit to the dimuon invariant mass distribution $m_{\mu \mu}$.
In order to enhance the sensitivity of the analysis, four bins in the c-BDT output (in increasing order of signal-over-background ratio) are defined in order to have constant signal efficiency equal to 18\% in each bin.
The fit is performed simultaneously in the four c-BDT bins. The first c-BDT bin, which has the lowest signal-over-background ratio, is dominated by the main backgrounds. It is introduced in the fit to improve the backgrounds modelling and reduce the systematic uncertainties related to them. 
The \Bsdmumu signals are parameterised by a double Gaussian function to take into account different resolutions in the dimuon invariant mass depending on the different regions of the ATLAS detector. The shape and the relative signal 
efficiencies are assumed to be the same in all c-BDT bins. 
The continuum background is described by a first order polynomial, while the background coming from \textit{SS} and \textit{SV} events is parameterised with an exponential function. 
These backgrounds are fluctuated independently in each c-BDT bin. Finally, the description of the peaking background is based on the same model used to describe the signal, with a constraint on the total yield of $2.9 \pm 2.0$ equally distributed in the c-BDT bins. 

The \BRof{\Bsdmumu} values are extracted through a simultaneous unbinned extended maximum-likelihood fit using the components written in Eq.~\ref{eq:ATLASMasterFormula} and the $N_{\Bsd}$ event yields extracted from the invariant mass fits just described.
The \BRof{\BuJpsimumuK} value is taken as the world average from the PDG~\cite{PDG2018}, while the hadronisation  probability ratio $\frac{f_{u}}{f_{d(s)}}= 0.256 \pm 0.013$ from the HFLAV average~\cite{HFLAV2016}.

The measurements are dominated by statistical uncertainties, with the most prominent sources of systematic uncertainty coming from: the fit uncertainties (where the largest contributors are the mass scale and the $b\to \mumu X$ background parameterisation), the $\frac{f_{u}}{f_{d(s)}}$ ratio (only for the BR($B^0_s$) measurement) and the reference channel yield. All systematic uncertainties are described in the likelihood as Gaussian constraints.

A Neyman construction~\cite{neyman} is employed to extract the 68.3\%, 95.5\% and 99.7\%
confidence intervals in the \BRof{\Bsmumu}--\BRof{\Bdmumu} plane. 
The likelihood function from the described Run-2 result is then combined with the likelihood function from the Run-1 result~\cite{Aaboud:2016ire}. The only common parameters in the combination are: the fitted \BRof{\Bsdmumu} and the external inputs (\BRof{\BuJpsimumuK}
and $\frac{f_{u}}{f_{d(s)}}$). All remaining nuisance parameters are treated as uncorrelated between the two results.

\begin{figure}[htb]                                                           \centering                                        \includegraphics[width=0.49\textwidth]{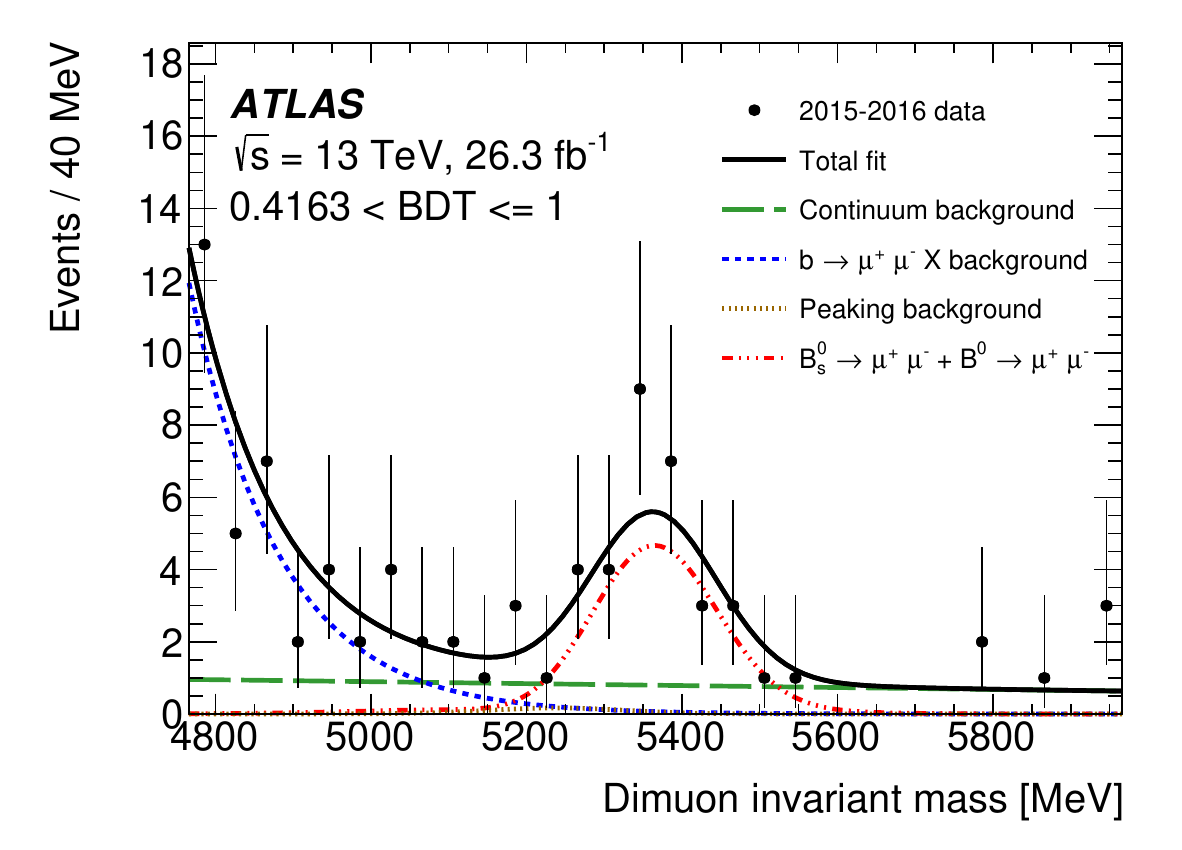}
  \includegraphics[width=0.49\textwidth]{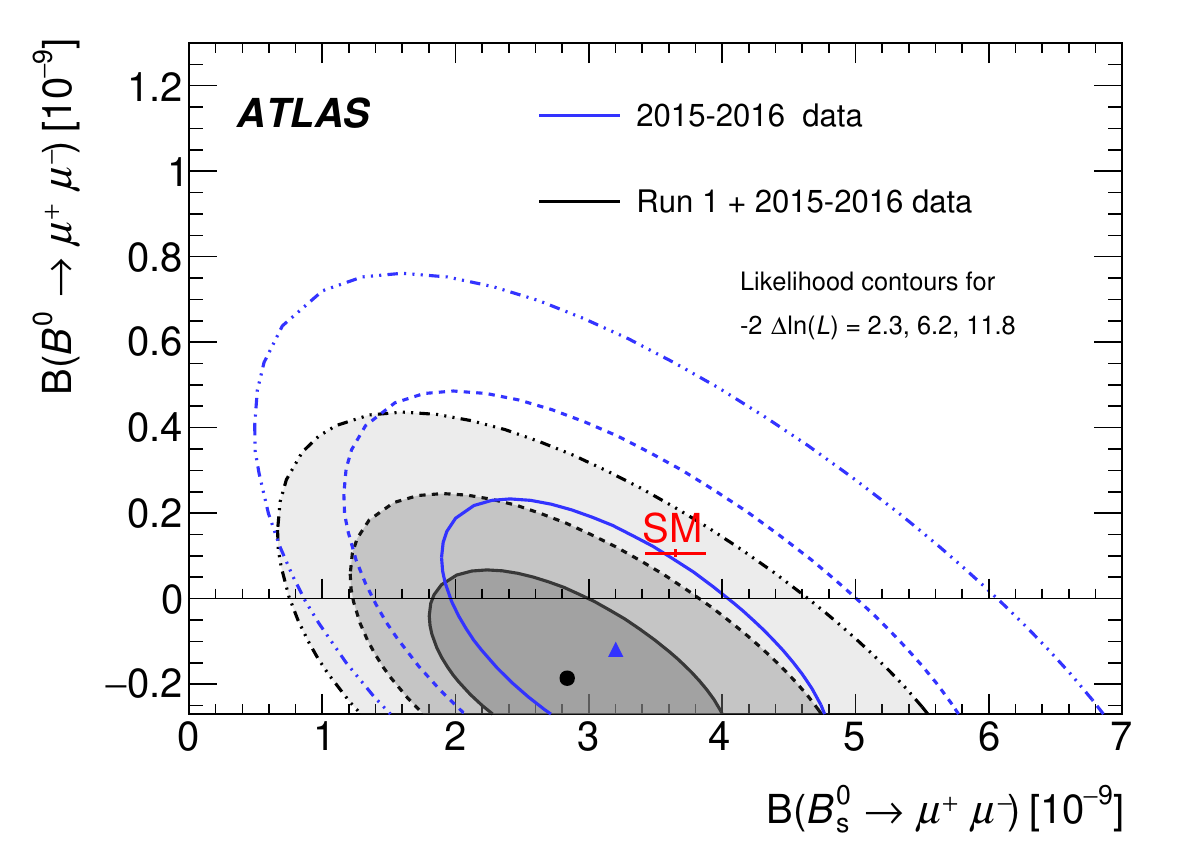}
  \caption{(Left): Dimuon invariant mass distributions in data, for the bin with the highest scores of BDT output. Superimposed is the result of the maximum-likelihood fit. The total fit is shown as a continuous line, with the dashed lines corresponding to the observed signal component, the $b \to \mu^+ \mu^-X$ background, and the continuum background. The signal components are grouped in one single curve, including both the \Bsmumu and the (negative) \Bdmumu component. The curve representing the peaking $B^0_{(s)} \to hh^\prime$ background lies very close to the horizontal axis. (Right): Likelihood contours for the combination of the Run 1 and 2015-2016 Run 2 results (shaded areas). The contours are obtained from the combined likelihoods of the two analyses, for values of $-2 \Delta \mathrm{ln}(L)$ equal to 2.3, 6.2 and 11.8. The empty contours represent the result from 2015-2016 Run 2 data alone. The SM prediction with uncertainties is also indicated. Figures from Ref.~\cite{Aaboud:2018mst}.}
  \label{fig:ATLAS_BR}                                        \end{figure}

The ATLAS results, obtained by combining 25~\invfb from Run1 and 26.1~\invfb from Run2 LHC campaigns, are~\cite{Aaboud:2018mst}:
\begin{eqnarray}
  \label{atlas-results}
  \BRof\Bsmumu & = & \left( 2.8 \,^{+0.8}_{-0.7} \right) \times 10^{-9} \quad ,\\
  \BRof\Bdmumu & = & \left( -1.9 \pm 1.6 \right) \times 10^{-10} \quad ,\nonumber
\end{eqnarray}
with a significance for the \Bsmumu signal of 4.6 standard deviations ($\sigma$).
The 95\,\% confidence level (CL) upper limit for the \Bdmumu signal is $\BRof\Bdmumu < 2.1 \times 10^{-10}$, as obtained with the Neyman procedure described in Ref.~\cite{neyman}.
Fig.~\ref{fig:ATLAS_BR} shows the dimuon invariant mass distribution in the highest-score BDT bin (left) and the likelihood contours in the \BRof{\Bsmumu} - \BRof{\Bdmumu} plane (right).\\

\subsection{The \Bsmumu effective lifetime measurement}

Using the same dataset and the same configurations for the event selection and the simulated samples, ATLAS has subsequently performed a measurement of the \Bsmumu effective lifetime \taumumu~\cite{lifetime_ATLAS}. As explained in Section \ref{sec:introduction}, the measurement of this quantity is complementary to the branching fraction measurement in the searches for NP phenomena.
The only difference between the two analyses lies in the different selection applied to the c-BDT output. A requirement on the c-BDT output to be larger than 0.365 is applied to the dataset, while all other requirements are the same as the BR analysis. The value of this requirement was selected after an optimisation procedure based on the maximisation of the $S/\sqrt{S+B}$ figure-of-merit.

The \Bsmumu effective lifetime is measured using a binned $\chi^2$ fit to the proper decay time distribution of the \Bsmumu signal component after the subtraction of the background. The proper decay time $\tilde{t}_{\mumu}$ is defined as $\tilde{t}_{\mumu} =\frac{L_{xy} m_{B_s^0}^\mathrm{PDG}}{p_\mathrm{T}^{B_s^0}}$, where $L_{xy}$ is the decay length projected along the reconstructed $B_s^0$ momentum in the transverse plane, $m_{B_s^0}^\mathrm{PDG}$ the world averaged mass of $B_s^0$ mesons from Ref.~\cite{PDG_2022} and $p_\mathrm{T}^{B_s^0}$ the magnitude of the candidate's reconstructed transverse momentum.
To extract \taumumu, three main steps have been completed:
\begin{itemize}
    \item A fit to the dimuon invariant mass, in the same range as for the BR analysis
    \item The extraction of the $\tilde{t}_{\mumu}$ distribution of the \Bsmumu component using the \textit{sPlot} technique~\cite{sPLOT}
    \item A binned $\chi^2$ fit to $\tilde{t}_{\mumu}$ distribution comparing Monte-Carlo simulated effective lifetime templates corresponding to different values of \taumumu.
\end{itemize}

In the first step, the dimuon invariant mass distribution, after all selection cuts described above, is fit using a five parameters model made of three Probability Density Functions (PDF): a double Gaussian to describe the \Bsmumu component, a linear function to describe the combinatorial (or \textit{continuum}) background component and an exponential function to describe the PRD component. Additional resonant and non-resonant backgrounds (such as $B \to h h'$, $B_c^{\pm}$ and semileptonic $B$ decays), as well as the \Bdmumu component, are neglected in this fit and considered as sources of systematic uncertainties whose impact on \taumumu~is evaluated through MC pseudo-experiments (as described later in the text). The fit yields 58$\pm$13 events in the \Bsmumu mass window.

The second step exploits the \textit{sPlot} statistical technique to extract the \Bsmumu signal proper-decay time component from the invariant mass fit. The signal proper-decay time distribution is background-subtracted by means of per-event weights computed using the result of the invariant mass fit described above.

\begin{figure}[htb]                                                           \centering                                                                                                               
  \includegraphics[width=0.49\textwidth]{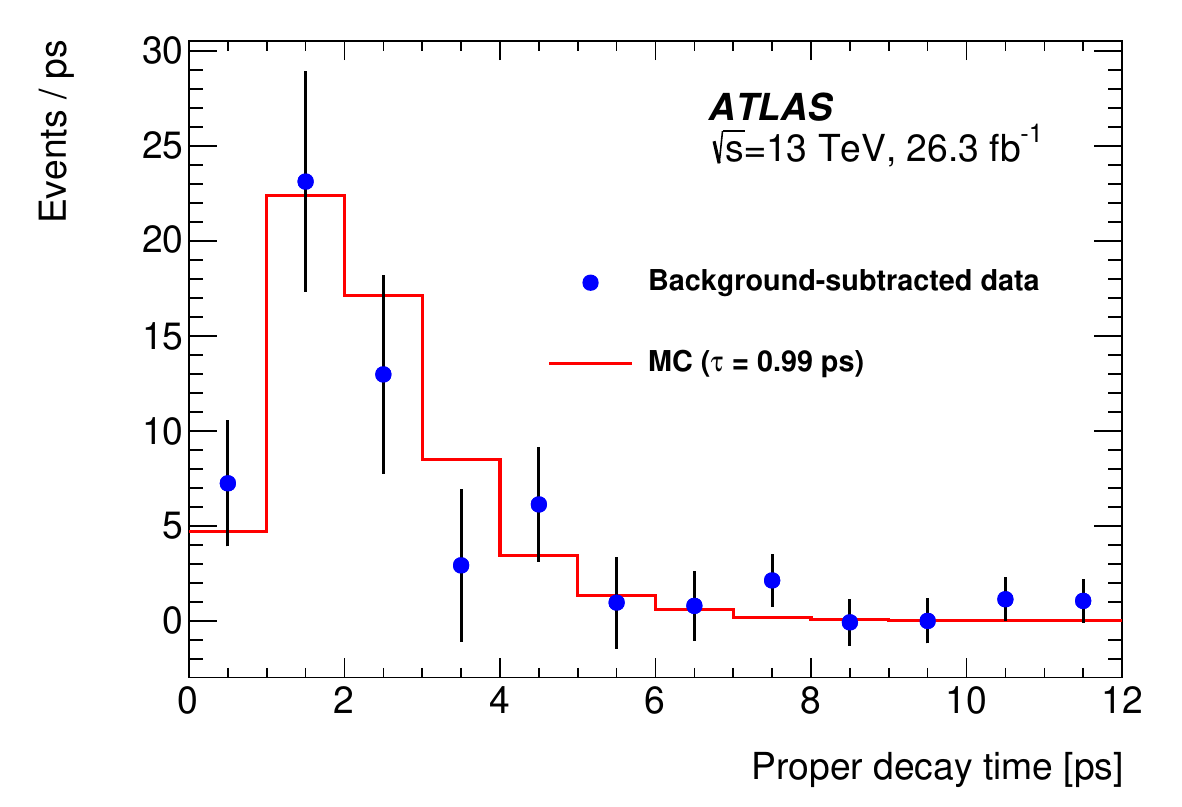}
  \includegraphics[width=0.49\textwidth]{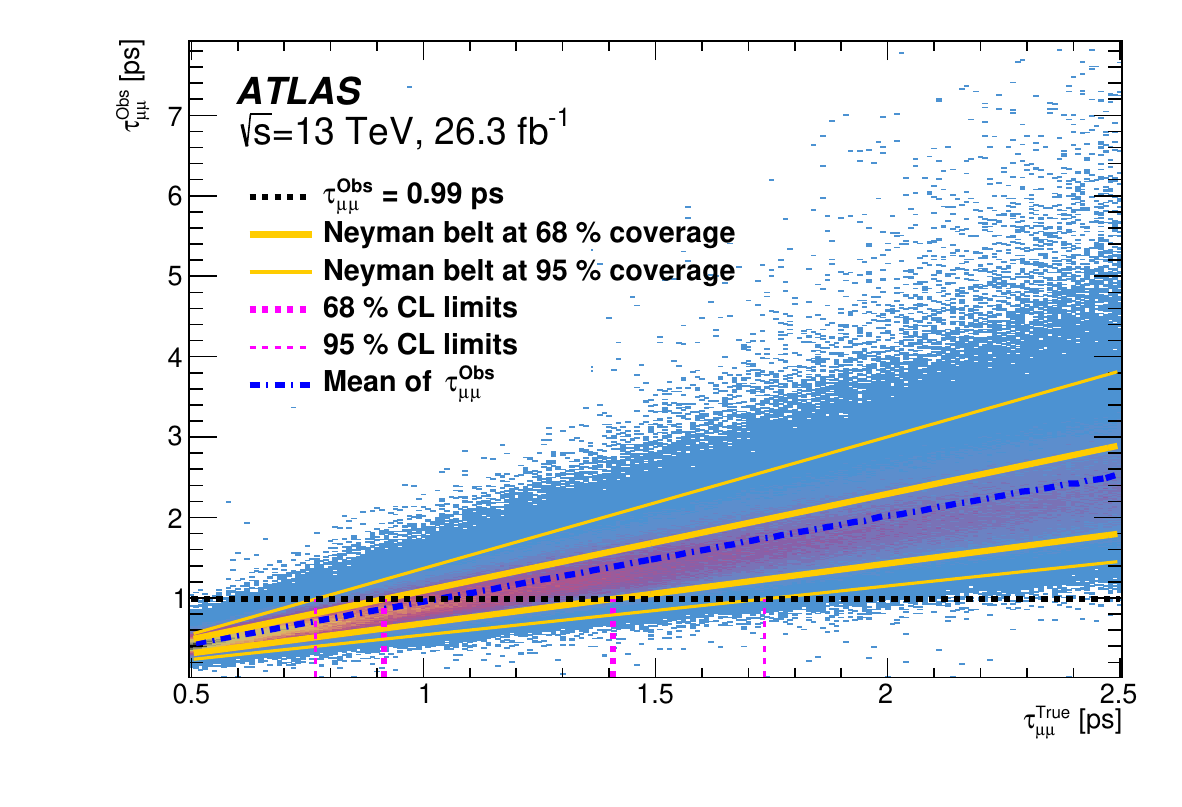}
  \caption{(Left): Signal proper decay time distribution extracted with the \textit{sPlot} background subtraction procedure applied to the \Bsmumu invariant mass fit. The superimposed signal MC template is the result of the lifetime fit procedure discussed in the text. The uncertainties on the data points are calculated as Poisson fluctuations on the MC yield prediction (continuous red histogram) in the corresponding bin.
  (Right): 68\% and 95\% CL bands obtained with a Neyman construction based on MC pseudo-experiments for the signal and background components. The yellow lines interpolate the band boundaries in order to smooth the effects of limited number of MC pseudo-experiments used. The dashed-dotted blue line corresponds to the average expected \taumumuobs\ value at a given \taumumu value used to generate MC pseudo-experiments. The horizontal dashed black line corresponds to the experimentally observed value of \taumumuobs=0.99 ps, yielding a 68\% CL band of [0.92,1.41] ps (thick vertical dashed purple lines) and a 95\% CL band of [0.77,1.73] ps (thin vertical dashed purple lines). The same construction at the \taumumuobs corresponding to \taumumu = 1.624 ps (the SM prediction) yields [1.44,2.26] ps as 68\% CL band. Figures from Ref.~\cite{lifetime_ATLAS}.}
  \label{fig:ATLAS_lifetime}                                              \end{figure}

The third and final step consists of a binned-$\chi^2$ fit to the proper-decay time distribution extracted in the previous step. This distribution is considered in the interval 0-12 ps and divided in twelve equal width bins. Pure signal proper-decay time simulated templates in the same interval and binning scheme corresponding to different values of \taumumu\ are generated, and a $\chi^2$-binned fit is performed with respect to background-subtracted data. The $\chi^2$ calculation takes both the statistical uncertainty on the weight-corrected MC and the Poissonian uncertainty in each data bin as expected from the predicted MC content for that bin into account. The template minimising the $\chi^2$ corresponds to an observed lifetime \taumumuobs\ of 1.07 ps. 
MC pseudo-experiments studies, generated for a lifetime of 1.624 ps (i.e. the SM predicted value) showed that the lifetime extraction procedure had a bias of 82 fs due to the low-statistics regime of the fit. This bias is found to be constant in the \Bs\ lifetime range considered in the analysis. Therefore the quoted value for \taumumuobs\ has been corrected for this effect.
The statistical uncertainty on \taumumuobs\ is instead extracted using a MC pseudo-experiments based Neyman construction, yielding to a value of $\taumumuobs = 0.99^{+0.42}_{-0.07}(\textrm{stat.})$ ps. Fig.~\ref{fig:ATLAS_lifetime} shows the signal proper decay time distribution extracted from data superimposed with the MC template minimising the $\chi^2$ distribution (left) and the MC pseudo-experiments based Neyman construction used to estimate the statistical uncertainty of the measurement (right). 

The dominant systematic uncertainties for this measurement are related to the data-MC discrepancies (134 fs evaluated in data by repeating, under the same statistical regime as the \Bsmumu signal case, the same fit procedure in the \BpmJpsiK channel), to the background mass and lifetime models (86 fs), to the fit dependence from the lifetime used in MC pseudo-experiments generation and the \Bs\ eigenstates admixture (15 fs evaluated by generating MC pseudo-experiments in the $\tau^{SM}_L$ - $\tau^{SM}_H$ lifetime interval) and the neglected resonant and non-resonant backgrounds (12 fs). The total systematic uncertainty is then obtained by summing in quadrature and symmetrising the impact on \taumumu\ of all single sources. 
This yields to an observed value of \taumumuobs\ of $0.99^{+0.42}_{-0.07} (\textrm{stat.}) \pm 0.17 (\textrm{syst.})$. The value is compatible with the SM prediction of 1.624 ps ($\ADeltaGamma = 1$) as well as with the other experimental results described in this article.\\

\section{Measurement of \boldmath \Bsmumu decay properties and search for \Bdmumu decay at CMS}
\label{sec:cms_BF}

The latest analysis by the CMS collaboration is based on the LHC Run-2 data collected in 2016--2018 at a center-of-mass energy of 13 TeV, corresponding to an integrated luminosity of 140 fb$^{-1}$~\cite{CMS-BPH-21-006}. The studies based on LHC Run-1 samples collected in 2011--2012 can be found in the earlier publications~\cite{CMS-BPH-16-004}. There is no attempt to combine the latest publication with the results from 2011–2012 data as the expected gain in sensitivity is modest. In this section the latest CMS measurement of \BRof\Bsmumu, the search for \Bdmumu decay, and the effective lifetime measurement using \Bsmumu events are discussed. 

\subsection{Measurement of \BRof\Bsmumu and search for \Bdmumu decay}

The characteristic signal \Bsdmumu comprises two muons originating from a single displaced vertex, isolated from other activities, with momentum aligned with the flight direction, and an invariant mass peaking at $M(\Bs)$ or $M(\Bd)$. The primary contributors to the background comprise combinatorial events, involving instances where the two muons originate from different heavy quarks, partially reconstructed semileptonic decays wherein both muons emanate from the same $B$ meson (with one of the muons from a misidentified charged hadron), and the background arising from peaking charmless two-body hadronic $B$ meson decays.

The data events were collected with a set of dimuon triggers for this study: the L1 trigger required two oppositely charged muons within the range of $|\eta|<1.5$, while at HLT the dimuon should form a secondary vertex and are required to be within specific mass ranges. The dimuon candidates are used to reconstruct $B$ mesons for the signal and normalization \BuToJPsiK and \BsToJPsiPhi channels. The selections are reserved to be as similar as possible for partial cancellation of systematic effects. Muons at offline analysis are required to have a high-quality track fit at tracker, a transverse momentum at least 4 GeV and $|\eta|<1.4$. To  suppress misidentified muons from charged pion and kaon decays, a multivariate-analysis(MVA)-based algorithm has been introduced. Extra kaons are required in the reconstruction for the normalization channels. A trajectory representing the $B$ candidate is built from the decay vertex and $B$ candidate's momentum, and is extrapolated to the closest point for each reconstructed primary vertex; the primary vertex with the smallest distance to the extrapolated point is selected for the analysis.

How to reduce the combinatorial and partially reconstructed backgrounds are the main challenges to the study. To enhance the analysis sensitivity, a dedicated MVA discriminator, combining various discriminating observables into a single score distribution ($d_{\rm MVA}$) using a boosted decision tree algorithm, is introduced. The inputs for $d_{\rm MVA}$ includes pointing angles, defined as the angles between the B momentum and the direction connecting the primary and secondary vertices, observables related to the secondary vertex such as quality of the vertex finding, and observables that are designed to identify nearby decay products in semi-leptonic decays of $b$ and $c$ hadrons. The $d_{\rm MVA}$ training is employed by an advanced gradient boosting algorithm, supported by the XGBoost library \cite{Chen:2016btl}. The training utilises a mix of \Bsmumu signal events and background events selected from the data sidebands. Following a fine-tuning of input observables to align the kinematics of \Bsmumu and \BuToJPsiK decays (considering variations in the uncertainties of the dimuon vertex position), the control decay \BuToJPsiK channel is employed to evaluate the performance of the $d_{\rm MVA}$ in data.

Charmless two-body decays \Bhhprime, like $B^0\to K^+\pi^-$ and $B^0_s\to K^+K^-$, can mimic the signal when both charged hadrons are misidentified as muons. The misidentification probabilities in data are calculated by utilizing $K^0_S\to\pi^+\pi^-$, $\phi(1020)\to K^+K^-$, and $\Lambda\to p\pi^-$ decays, restricting the decay distance of $K^0_S$ and $\Lambda$ candidates to align with the lifetime of the $B$ meson. Misidentification of pions and kaons primarily originates from their decays into muons. An agreement between the observed data and simulations is observed for both pions and kaons. The proton misidentifying rate is much smaller hence the contributions from the associated processes are totally negligible. After stringent multivariate-based muon identification requirement, the charmless two-body backgrounds reduce to a negligible level.

Because of the limited precision in measuring the b-quark production cross section at the LHC, directly determining the branching fraction (\BRof{\Bsdmumu}) could introduce significant uncertainty. As a common practice, the signal branching fraction is assessed by normalizing it to the \BuToJPsiK decay channel. In addition the \BsToJPsiPhi decays, with $J/\psi \to \mu \mu$ and $\phi \to KK$, are considered as a cross-check, and might become more precise if the $\BRof{B^0_s \to J/\psi \phi(1020)}$ is further improved by future B-factory studies. Another advantage of measuring branching fractions in a relative manner is the potential cancellation of systematic uncertainties common in the selection of the signal and normalization channels. The exact formulae for the 
\Bsdmumu branching fractions are similar to those used in Eq.\ref{eq:ATLASMasterFormula}:
\begin{align}
\BRof{\Bsmumu}&=\BRof{B^+\to J/\psi K^+}{ N_{\Bsmumu} \over N_{B^+\to J/\psi K^+}}{  \epsilon_{B^+\to J/\psi K^+} \over \epsilon_{\Bsmumu} } {f_u\over f_s},\\
\BRof{\Bsmumu}&=\BRof{\BsToJPsiPhi}{ N_{\Bsmumu} \over N_{B_s^0 \to J/\psi \phi}}{  \epsilon_{B^+\to J/\psi \phi} \over \epsilon_{\Bsmumu} },\\
\BRof{\Bdmumu}&=\BRof{B^+\to J/\psi K^+}{ N_{\Bdmumu} \over N_{B^+\to J/\psi K^+}}{  \epsilon_{B^+\to J/\psi K^+} \over \epsilon_{\Bdmumu} } {f_u\over f_d},
\end{align}
where the yields and the selection efficiencies for each processes are denoted by $N_X$ and $\epsilon_X$ ($X = \Bsmumu$, $\Bdmumu$, $B^+\to J/\psi K^+$, or \BsToJPsiPhi). The  production fractions for $B^+$, $B^0$, and $B^0_s$ mesons are represented by $f_u$, $f_d$, and $f_s$.  The ratio $\frac{f_{u}}{f_{d}}$ is 
set to unity due to isospin symmetry, while the ratio $\frac{f_{u}}{f_{s}}$, together with $\BRof{B^+\to J/\psi K^+}$ and $\BRof{B^0_s\to J/\psi \phi}$, are external inputs. 

The results are obtained with a simultaneous unbinned maximum likelihood fits in multiple categories. For the measurement of branching fractions, a two-dimensional fit using the dimuon invariant mass and its uncertainty as observables is introduced. The events are categorized according to data-taking period, signal purity based on $d_{\rm MVA}$, and $|\eta|$ of the most-forward muon.  The likelihood function include five components: $\Bsmumu$ and $\Bdmumu$ signals, 
semileptonic background, peaking two-body decays, and the combinatorial events. The signal components are represented using Crystal Ball functions for the dimuon mass. The width of these Crystal Ball functions is parameterized based on the per-event mass resolution. To model the mass resolution, a kernel estimation approach is employed, utilizing Gaussian kernels. The semileptonic background is modeled by a Gaussian with free parameters in the fit to the data, while the peaking background is modeled by a sum of Gaussian and Crystal Ball functions with the shape parameters determined from simulated events. The yields of the semileptonic and peaking background components are first derived and then included in the fit with uncertainties from the hadron to muon misidentifying rate as constrained nuisance parameters. The combinatorial background is modeled by a linear function with yields and slope free to vary in the fits. 

For the branching fraction measurements the experimental uncertainties include signal efficiency corrections due to mismodeling of $d_{\rm MVA}$, the charged kaon efficiency in the normalization channels, trigger efficiencies, and fitting bias, while the rest of uncertainties are smaller than 1\%. The mismodeling of the $d_{\rm MVA}$ distribution has been investigated through two distinct studies with $B^+\to J/\psi K^+$ events. In the first study, a direct comparison is conducted between background-subtracted data, with the \textit{sPlot} technique~\cite{Pivk:2004ty} on the $B^+\to J/\psi K^+$ invariant mass distribution, and the simulated distributions. The second study involves reweighting of the simulated samples to align with the data distributions, employing the XGBoost tool. The disparity between the two studies is quantified as a systematic uncertainty. The systematic uncertainty arising from the selection of background models is derived through pseudo-experiments, incorporating variations in the fit. The uncertainties in the input branching fractions of the normalization channels and the $\frac{f_{u}}{f_{d(s)}}$ ratio, are implemented as external uncertainties. 

\begin{figure}[H]
\centering
\includegraphics[width=0.49\textwidth]{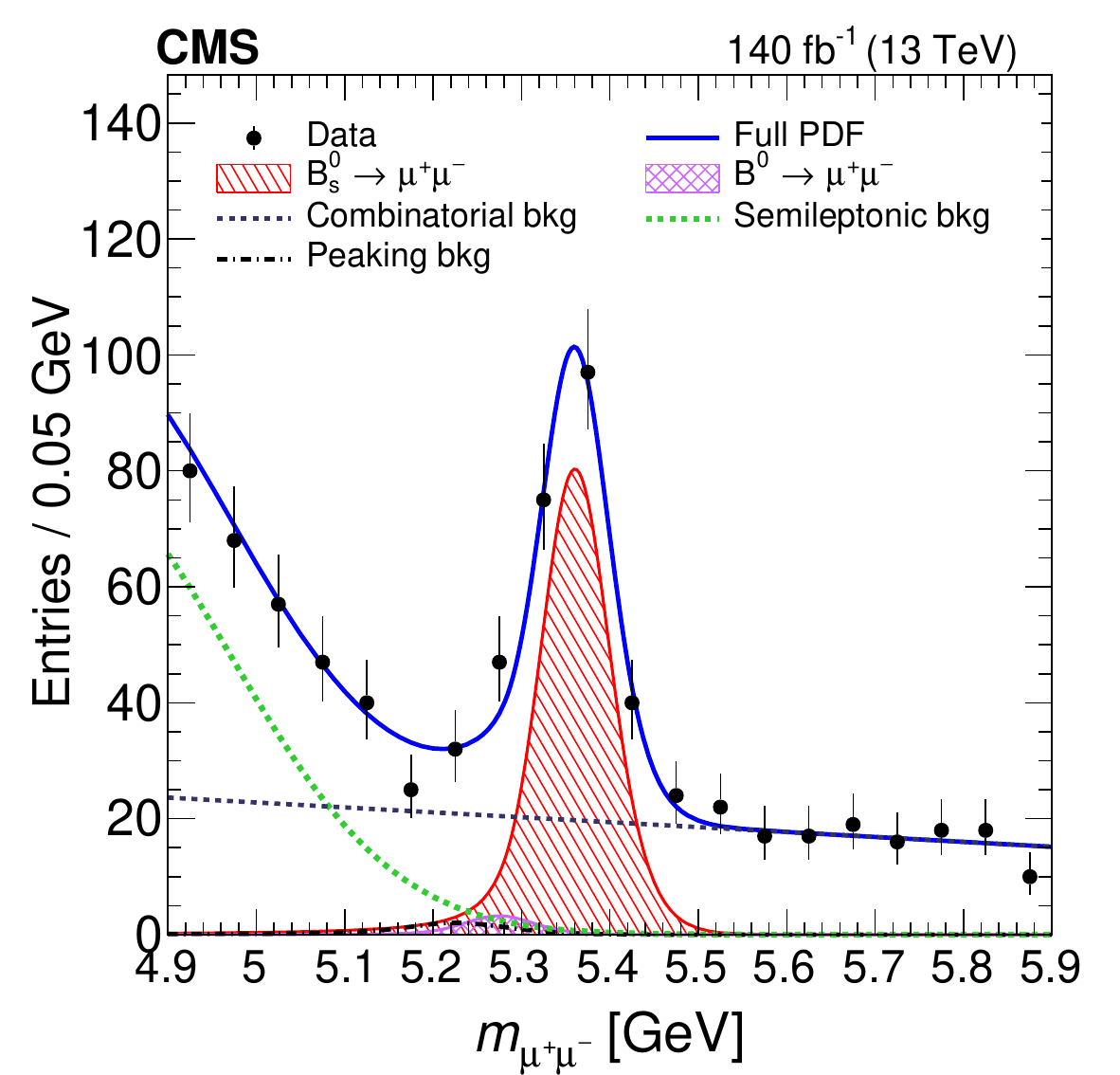}
\includegraphics[width=0.49\textwidth]{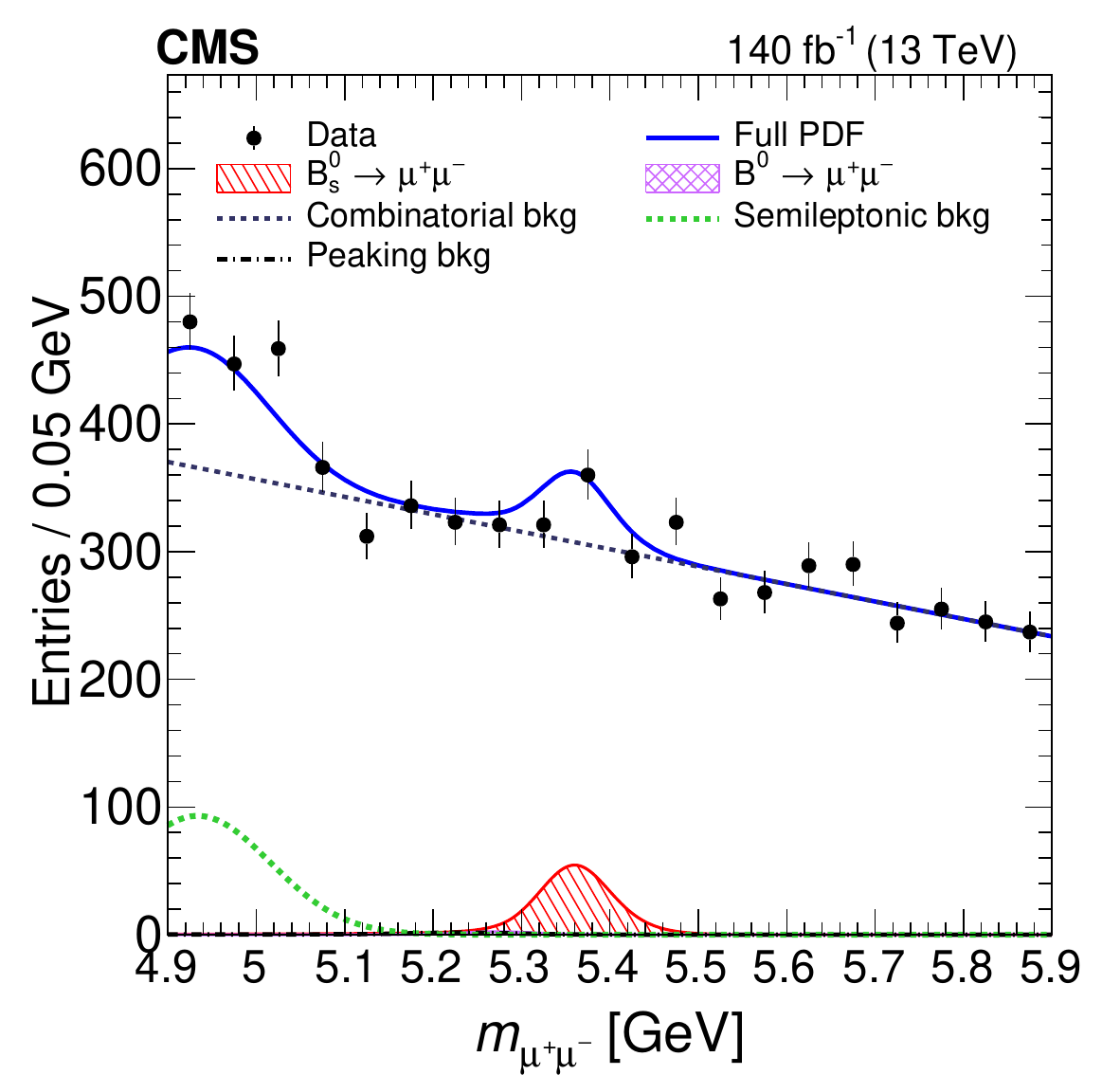}
\caption{The dimuon invariant mass distributions for the candidates with $d_{\rm MVA}>0.99$ (left) and $0.99 >d_{\rm MVA}> 0.90$ (right) categories. The solid blue curves are the projections of fit model, while the individual components of the fit are also presented.
Figures from Ref.~\cite{CMS-BPH-21-006}.
\label{fig:cms_mass_proj}}
\end{figure}

\begin{figure}[H]
\begin{center}
\includegraphics[width=8 cm]{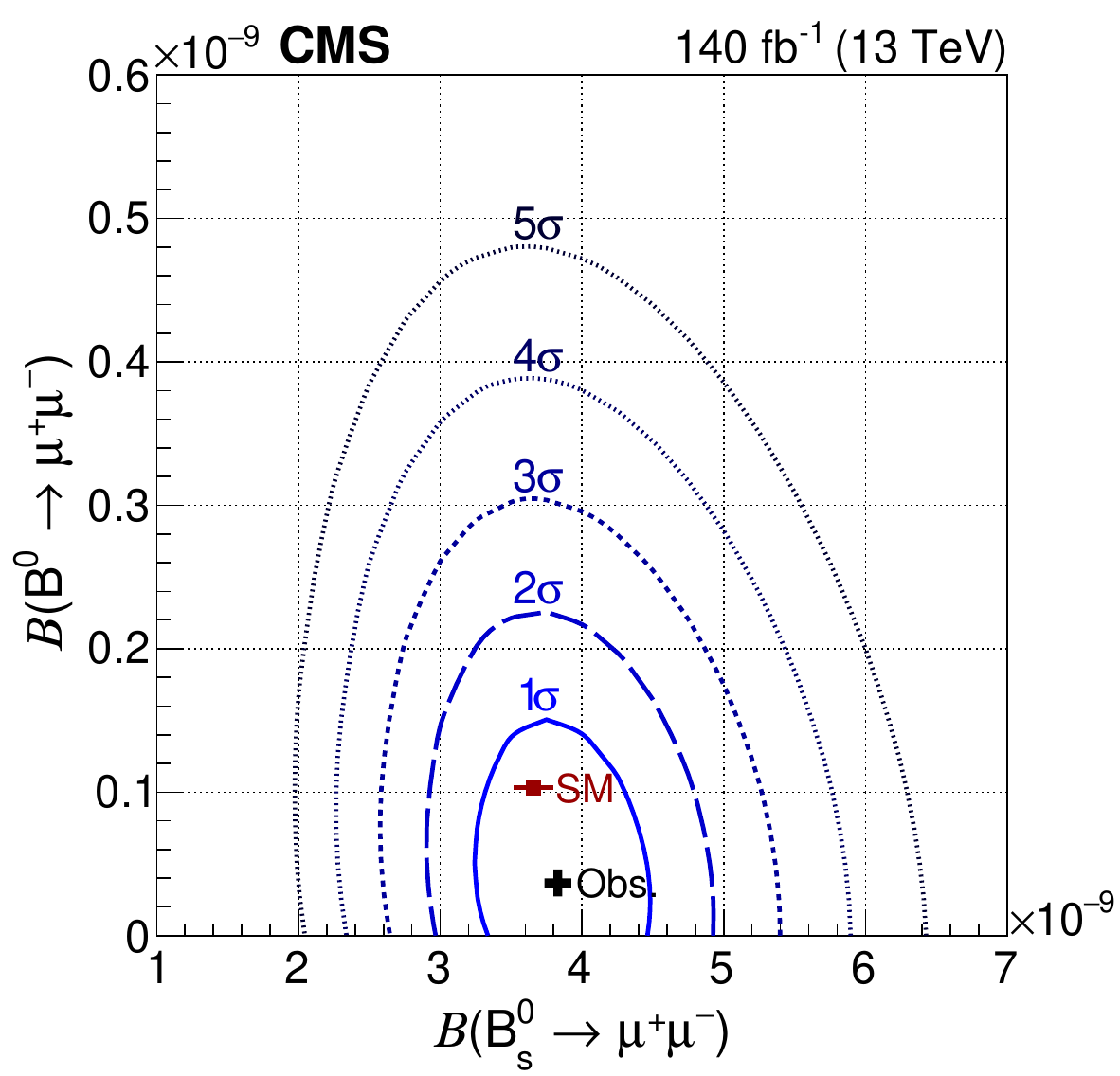}
\caption{The profile likelihood (left) in $\BRof{\Bsmumu}$ versus $\BRof{\Bdmumu}$ 2D plane, while the contours enclose the regions with $1\sigma$-$5\sigma$ coverage. 
Figure from Ref.~\cite{CMS-BPH-21-006}.
\label{fig:cms_2dcontours}}
\end{center}
\end{figure}

The resulting branching fractions for  $\Bsmumu$ and $\Bdmumu$ are:
\begin{align}
\BRof{\Bsmumu}&=\left[ 3.83^{+0.38}_{-0.36}\,{\rm (stat)}^{+0.19}_{-0.16}\,{\rm (syst)}^{+0.14}_{-0.13}\,{(f_s/f_u)} \right]\times 10^{-9},\\
\BRof{\Bdmumu}&=\left[ 0.37^{+0.75}_{-0.67}\,{\rm (stat)}^{+0.08}_{-0.09}\,{\rm (syst)} \right]\times 10^{-10}.
\end{align}
The results incorporate external inputs, specifically $\BRof{B^+\to J/\psi K^+} = (1.020\pm0.019)\times 10^{-3}$, $\BRof{J/\psi \to \mu^+\mu^-} = (5.961\pm0.033)\times 10^{-2}$, and $f_s/f_u = 0.231\pm0.008$. The input $f_s/f_u$ value is derived from the $p_T$-dependent measurement by LHCb~\cite{LHCb:2021qbv} and the $p_T$
distribution observed in this measurement. Figure~\ref{fig:cms_mass_proj} shows the dimuon invariant mass distributions 
from the categories with different signal purity; the results of the fit are superimposed. The profile likelihood contours enclose the regions with different coverage are shown in Figure~\ref{fig:cms_2dcontours}.
Alternatively the $\Bsmumu$ branching fraction is measured using the $B^0_s\to J/\psi \phi$ decays as the normalization, which leads to 
\begin{align}
\BRof{\Bsmumu}&=\left[ 4.02^{+0.40}_{-0.38}\,{\rm (stat)}^{+0.28}_{-0.23}\,{\rm (syst)}^{+0.18}_{-0.15}\,{(\mathcal{B})} \right]\times 10^{-9},
\end{align}
where the last uncertainty arises from the uncertainty in the $B^0_s\to J/\psi \phi$ branching fraction ($\BRof{B^0_s\to J/\psi \phi} = (1.04\pm0.040)\times 10^{-3}$). 
The lifetime of the $B^0_s$ meson has a significant impact on the $\Bsmumu$ branching fraction too; a scaling factor on the resulting branching fraction ($1.577-0.358 \cdot \tau_{B^0_s}$, where $\tau_{B^0_s}$ is $B^0_s$ lifetime in ps) for alternative lifetime hypotheses other than the SM value (1.61 ps) is provided. 
The upper limit on the $\Bdmumu$ decay is calculated to be 
$\BRof{\Bdmumu}<1.9\times 10^{-10}$ at 95\% confidence level, using the CL$_s$ method~\cite{Read:2002hq}.

\unskip

\unskip

\subsection{Measurement of \Bsmumu Effective Lifetime}

The \Bsmumu effective lifetime \taumumu is extracted with a unbinned maximum likelihood in three-dimensions including 
dimuon invariant mass, decay time, and decay time uncertainty. The decay time $t_{\mumu}$, which is calculated for each event, is defined by the product of the flight length and the invariant mass of the $B$ candidate, divided by the magnitude of the $B$ candidate momentum. The events are also categorized in the data-taking period, purity based on $d_{\rm MVA}$,  and the pseudorapidity of the most forward muon. The dimuon invariant mass distribution is modeled with the same functions introduced for the branching fraction measurements, while the decay time distribution for signal events is modeled by an exponential function convoluted with the decay time resolution function. The decay time resolution function is parameterized with the measured decay time uncertainty. The acceptance as a function of the decay time is obtained from simulated events and corrected with the \BuToJPsiK events from data. The decay time distribution for combinatorial background decays is obtained from high-mass sideband events. The decay time uncertainty models used in the fit are obtained from simulation samples and mass sideband data as well. 

The systematic uncertainties in the lifetime measurement are mostly driven by the correlations between the $d_{\rm MVA}$ and the decay time, as 
the key input variables for the $d_{\rm MVA}$ classifier: the pointing angle of $B$ candidate and its associated uncertainty are strongly correlated with the decay time observable. Any mismodeling in the simulation results in significant impacts on the decay time distribution. A correction as a ratio of the decay time distributions for different  $d_{\rm MVA}$ requirements is derived from \BuToJPsiK events. This method introduced a bias up to 0.1~ps for the data recorded in 2016, and reduced in the later data sets. The possible bias arises in fitting and modeling is also tested with \BuToJPsiK events, but with a relaxed selection criterion. Other systematic uncertainties are minor, estimated to be smaller than 0.01~ps.

\begin{figure}[H]
\begin{center}
\includegraphics[width=8 cm]{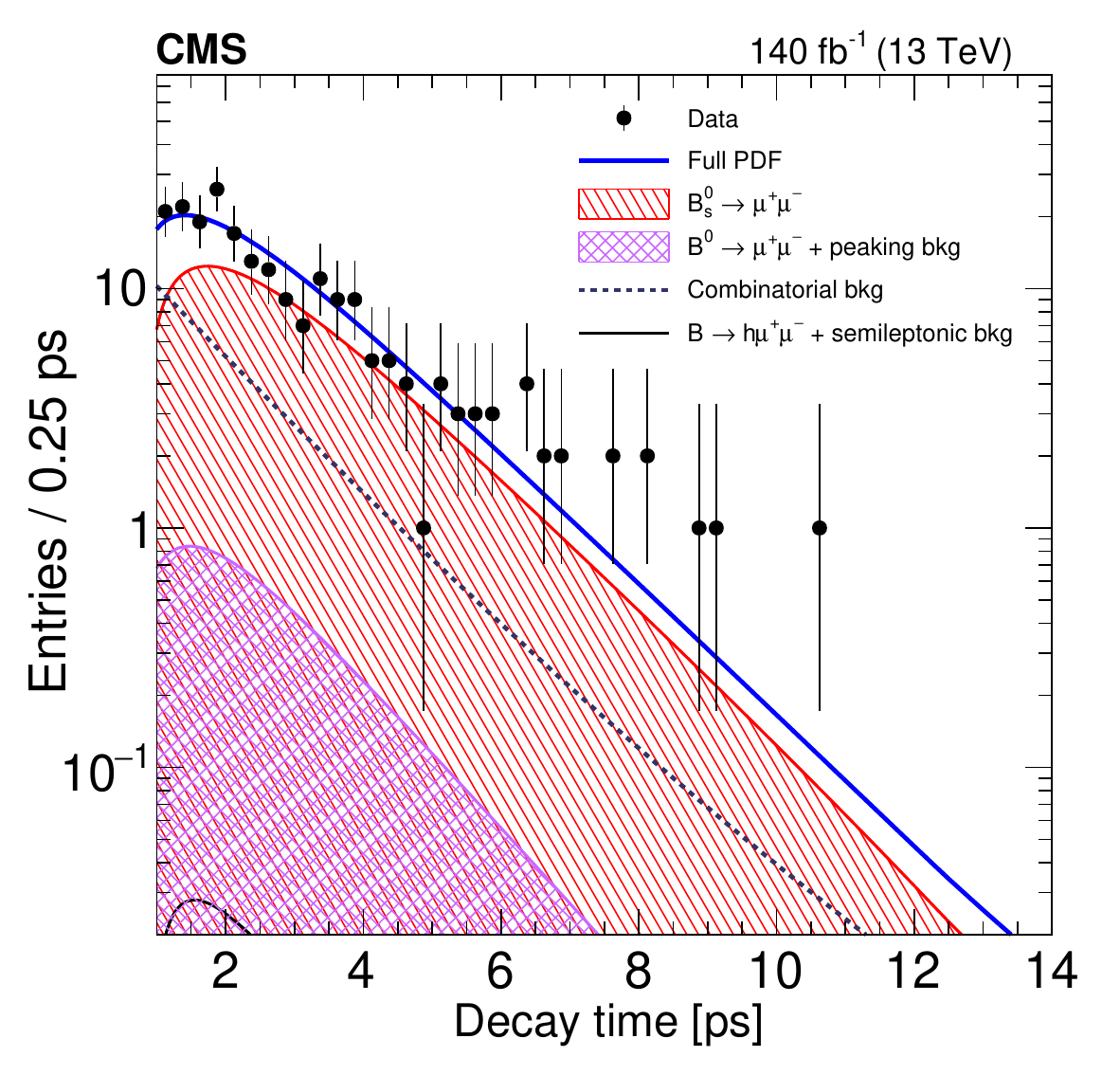}
\caption{The proper decay time distribution for the candidates 
in the region of $5.28 < m_{\mu^+\mu^-} < 5.48$~GeV, with the result of the fit superimposed. The solid blue curve is the sum of all fit component, while the shaded areas are the background components. 
Figure from Ref.~\cite{CMS-BPH-21-006}.
\label{fig:cms_decaytime_proj}}
\end{center}
\end{figure}

The resulting effective lifetime for $\Bsmumu$ events is:
\begin{align}
\taumumu = 1.83^{+0.23}_{-0.20}\,{\rm (stat)}^{+0.04}_{-0.04}\,{\rm (syst)}~{\rm ps},
\end{align}
which is consistent with the SM prediction and the other experimental results described in this article.  The decay time distribution for the candidates 
in the region of $5.28 < m_{\mu^+\mu^-} < 5.48$~GeV is shown in Fig.~\ref{fig:cms_decaytime_proj}.

\section{Analysis of \boldmath\Bsdmumu decays with LHCb}
\label{sec:LHCb_measurements}

The most recent analysis of \Bsdmumu with the LHCb experiment~\cite{LHCb:2021awg,LHCb:2021vsc} was performed with the full $pp$-collision data collected in the LHC Run 1 and Run 2 campaigns.
The total integrated luminosity corresponds to $1\,\text{fb}^{-1}$ at $\sqrt{s} = 7\,\tev$, $2\,\text{fb}^{-1}$ at $\sqrt{s} = 8\,\tev$ and $6\,\text{fb}^{-1}$ at $\sqrt{s} = 13\,\tev$.

In total the analysis comprises the search and branching fraction measurements of the decays \Bsmumu, \Bdmumu and \Bsmumugamma with initial state radiation (\Bsmumugamma was only investigated in the region $m(\mumu)>4.9\,\gevcc$), as well as the measurement of the effective lifetime of the \Bsmumu decay.
A precise branching fraction measurement is achieved by normalising the signal decay with two high-statistics decay modes, \BdKpi and \bujpsik with $\decay{\jpsi}{\mumu}$, similarly to what is done by ATLAS and CMS Collaborations and shown in Eq.\ref{eq:ATLASMasterFormula} for the \bujpsik channel. The decay modes \BdKpi and \BsKK are used as control modes for the effective lifetime measurement as well.

Dominant background processes mimicking the signal on the one hand arise from random combinations of two muons from two different \bquark-hadron decays in the same event. On the other hand they can come from \bquark-hadron decays where one or more final state particles have been wrongly identified as a muon. Furthermore, \bquark-hadron decays wher part of the decay products have not been reconstructed can constitute a background. The selection of the signal decays largely inherits from previous analyses of a subset of the data~\cite{LHCb-PAPER-2017-001} and targets particularly the selecting of \Bsdmumu decays over aforementioned backgrounds, whereas the measurement of \Bsmumugamma is a byproduct of the analysis. The LHCb detector, as used to collect the above mentioned data,  employed a two-staged online selection. Firstly, events are selected by a hardware trigger that requires at least one muon with a high transverse momentum. Secondly, a two-staged software trigger is applied, which performs a full event reconstruction. In the software trigger, events fulfilling minimum requirements on the muon momentum and its impact parameter, are kept. Also events are kept where these requirements are met by non-signal candidates to maximise signal efficiency.

In the offline selection, candidate \Bsdmumu decays are selected by combining two well reconstructed oppositely charged particles identified as muons~\cite{Archilli:2013npa} with a transverse momentum in the range of $0.25<\pt<40\,\gevc$. The common vertex is required to have a good vertex fit quality and be clearly separated from the associated $pp$-collision vertex. The resulting \Bsd candidate is required to have a transverse momentum greater than $0.5\,\gevc$. Candidates in the full instrumented pseudorapidity region $2<\eta<5$ are retained for analysis. A preliminary selection based on a Boosted Decision Tree (BDT) is applied to remove a large fraction of combinatorial background while maintaining a high signal efficiency. The BDT is trained with variables related to the decay topology of two particles originating from a vertex displaced with respect to the primary vertex.
A highly efficient veto on the combination of a signal muon with another particle in the event identified as muon that result in a dimuon mass close to the \jpsi mass allows to effectively remove \decay{\Bc}{\jpsi\mup\nu} decays.
A selection on a combination of particle identification algorithms is performed and tuned to maximise the \Bdmumu significance ($\hadron=\kaon,\pion$), suppressing \Bhh and $\Lb\to\proton\mun\nu$ decays.
The final selection is performed on a second BDT, called in the following s-BDT. This s-BDT includes, apart from variables related to the decay topology, notably isolation classifiers - specifically developed for this analysis - that inspect the closeness of the signal muon tracks to other tracks in the event that are either reconstructed in all tracking detector stations or only in the detector closest to the collision region.
The \Bsdmumu yields are measured by fitting the dimuon invariant mass distribution in bins of this final selection s-BDT, discarding only the lowest bin (that corresponds to about $25\,\%$ of the signal) in order to maximise the signal sensitivity.
The samples of \BdKpi and \bujpsik are selected in a similar way except for trigger and particle identification criteria for the \BdKpi mode and removing the \jpsi veto. For \BdKpi, the muon identification criteria are replaced by hadron identification and a trigger selection independent of the candidate is required to achieve an unbiased selection.

\subsection{Measurement of the branching fractions of \Bsdmumu and \Bsmumugamma}
In order to achieve unbiased branching fraction estimates, efficiencies are calculated either on corrected simulation or directly on data.
Importantly, the fractions of the s-BDT bins are determined from \Bsdmumu simulation, where the \Bsd quantities and the number of tracks in the event are reweighted from data-simulation comparisons in high-statistics \bujpsik and \BsToJPsiPhi samples. The resulting corrected s-BDT fractions are then independently cross-checked with \BdKpi data samples, corrected by the different trigger and particle identification response.
Measuring the branching fractions relative to two modes, \BdKpi and \bujpsik, allows for a stringent cross check of the efficiencies by calculating the ratio between the estimated branching fractions of the two and comparing it to the ratio of the published branching fractions~\cite{PDG2020}.
An excellent agreement is found.

The invariant mass shape of signal \Bsdmumu decays is described with two-sided Crystal Ball functions~\cite{Skwarnicki:1986xj}, where the mean of the Gaussian core is calibrated from \BsKK and \BdKpi data samples. The mass resolution of about $22\,\mevcc$ is determined from the interpolation of the measured resolutions of charmonium and bottomonium resonances. The tail parameters are estimated from simulation. Small differences in the resolution and the tail parameters are found to appear across the s-BDT bins and are accounted for in the final fit.

Exclusive background decays remaining in the fully selected samples have been carefully studied with simulation, calibrated in data.
A large focus in the most recent analysis is laid on the correct estimation of the misidentification of charged hadrons as muons.
Decays of the form \Bhhprime ($h=\kaon,\pion$) with both charged hadrons misidentified create a peaking structure very close to the \Bdmumu peak and therefore form the most relevant remaining background component. Misidentification occurs in the detector dominantly because the hadrons decay in-flight into muons.
The hadron misidentification rate is estimated with a dedicated procedure using \decay{\Dz}{\Km\pip} from \decay{\Dstarp}{\Dz\pip} decays from simulation and data. This procedure takes explicitly into account that the \decay{\Dz}{\Km\pip} invariant mass shape deforms significantly with hadrons decaying in-flight.
As additional cross check, the misidentification rate is investigated from \BdKpi data samples by determining the \BdKpi yield in \pion\muon, \kaon\muon and \pion\kaon mass distributions.

\begin{figure}
\begin{center}
\includegraphics[width=0.49\textwidth]{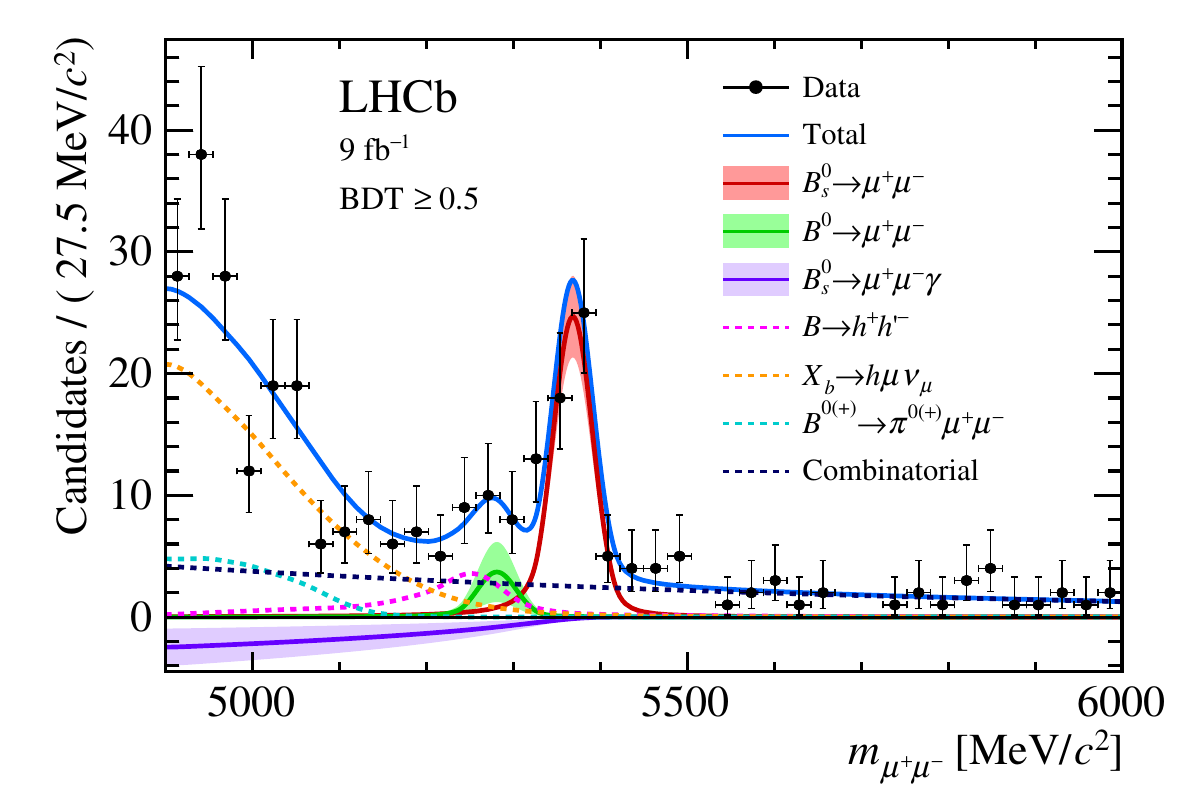}
\includegraphics[width=0.49\textwidth]{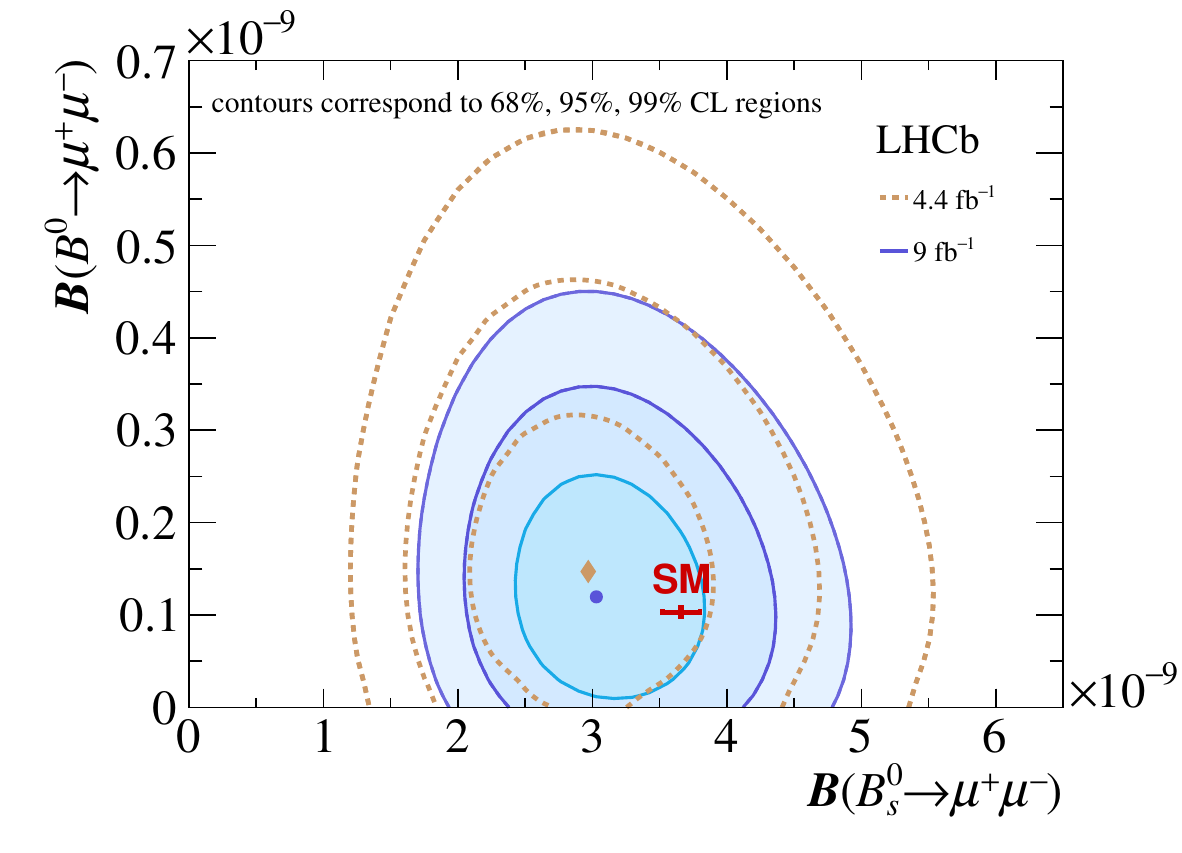}
\caption{Left: mass distribution of the selected \Bsdmumu candidates (black dots) with s-BDT\,$> 0.5$. The result of the fit is overlaid and the different components are detailed: \Bsmumu (red solid line), \Bdmumu (green solid line), \Bsmumugamma (violet solid line), combinatorial background (blue dashed line), \Bhh (magenta dashed line), \decay{\Bd}{\pim\mup\neum}, \decay{\Bs}{\Km\mup\neum}, \decay{\Bcp}{\jpsi\mup\neum} and \decay{\Lb}{\proton\mun\neum} (orange dashed line), and $\decay{B^{0(+)}}{\pi^{0(+)}\mup\mun}$ (cyan dashed line). The solid bands around the signal shapes represent the variation of the branching fractions by their total uncertainty. Right: two-dimensional profile likelihood of the branching fractions for the \Bsdmumu decays. The measured central values of the branching fractions are indicated with a blue dot. The profile likelihood contours for $68\,\%$, $95\,\%$ and $99\,\%$ CL regions of the result are shown as blue contours, while the brown contours indicate the previous measurement~\cite{LHCb-PAPER-2017-001} and the red cross shows the SM prediction. Figures from Ref.~\cite{LHCb:2021vsc}.}
\label{fig:BF_fit_LHCb}
\end{center}
\end{figure}

A summary of the final mass fit to obtain the signal branching fractions is displayed in Fig.~\ref{fig:BF_fit_LHCb}.
A precise measurement of the \Bsmumu branching fraction is obtained to be
\begin{align}
	\BRof{\Bsmumu}=\left(3.09^{+0.46}_{-0.43} {}^{+0.15}_{-0.11}\right)\times10^{-9},
\end{align}
where the first uncertainties are of statistical and the second of systematic nature.
The systematic uncertainties are dominated by the knowledge of the ratio of fragmentation fractions $\frac{f_{u}}{f_{d(s)}}$ of \Bs and \Bd mesons which enters the normalisation equation because the decay is measured relative to \Bd and \Bu decays.
The \Bdmumu and \Bsmumugamma decays are not observed and consequently upper limits on their branching fractions are set to
\begin{align}
	\mathcal{B}(\Bdmumu)&<1.2\times10^{-10}\text{, and}\\
	\mathcal{B}(\Bsmumugamma&)<2.0\times10^{-9}
\end{align}
at $95\,\%$ CL, respectively.
Similarly, an upper limit on the branching fraction ratio $\mathcal{R}$ was determined at $95\,\%$ CL to
\begin{align}
	\mathcal{R}<0.095
\end{align}
These values include systematic uncertainties, which are dominated by the knowledge of the background components that include misidentified hadrons.
A correlation of $11\,\%$ is observed between the measurement of the \Bdmumu and \Bsmumu components.

\subsection{Measurement of the effective lifetime of the \Bsmumu decay}
\begin{figure}
\begin{center}
\includegraphics[width=0.49\textwidth]{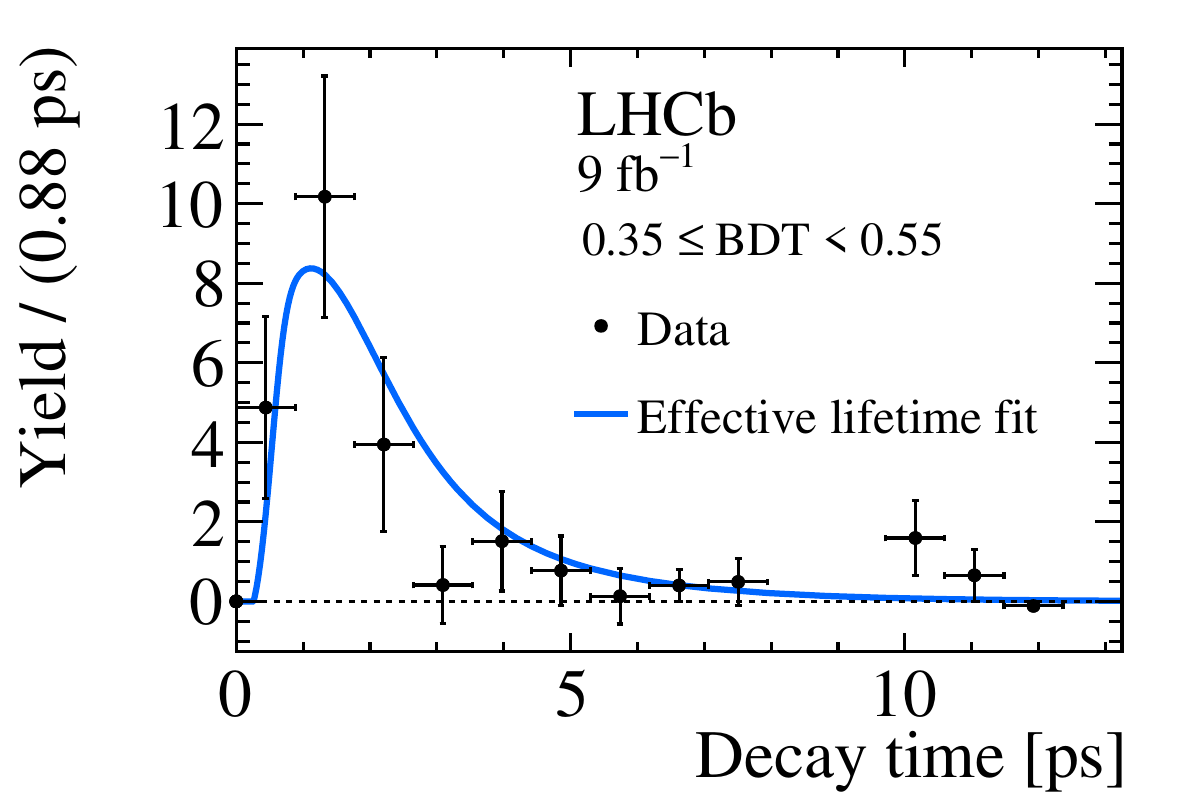}
\includegraphics[width=0.49\textwidth]{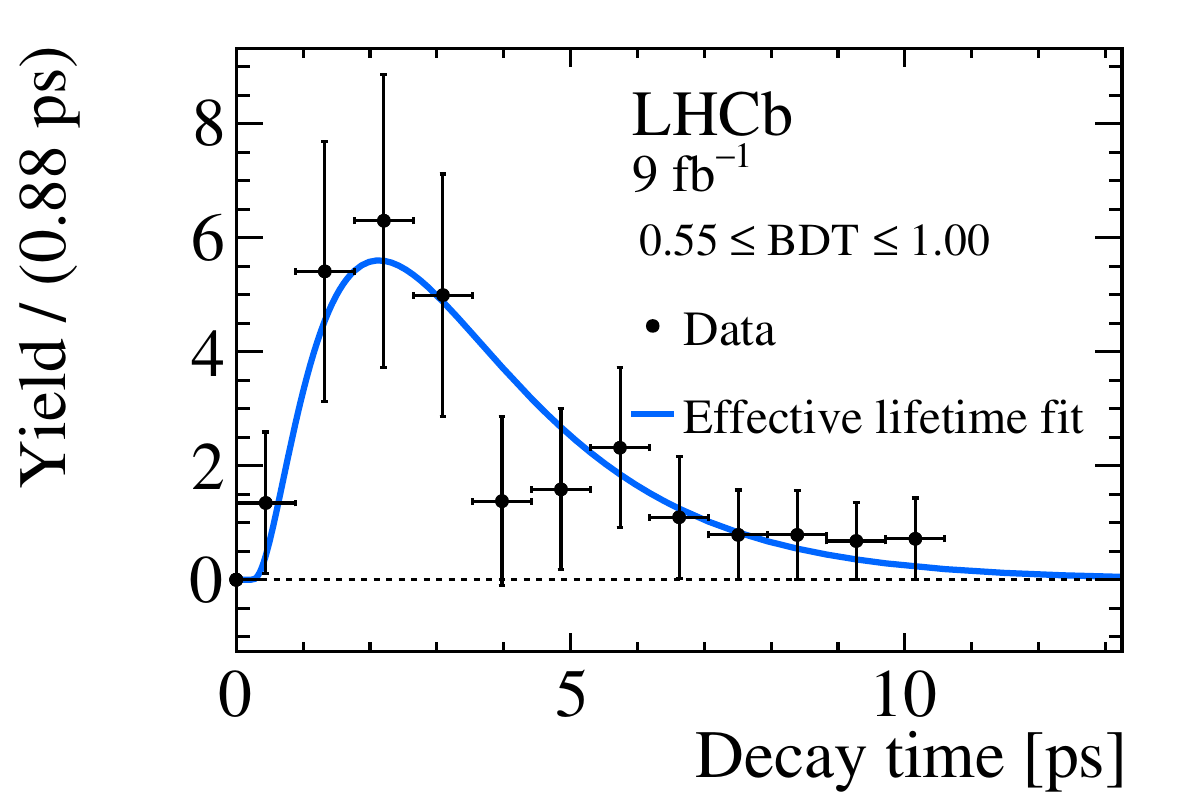}
\caption{The background-subtracted decay-time distributions with the fit model used to determine the \Bsmumu effective lifetime superimposed. The distributions in the low and high BDT regions are shown in the left and right plot, respectively. Figures from Ref.~\cite{LHCb:2021vsc}}
\label{fig:lifetime_fit_LHCb}
\end{center}
\end{figure}

The effective lifetime of the \Bsmumu decay has been measured on the same sample with a slightly different selection.
Since there is effectively no background from hadron-muon misidentification in the \Bsmumu mass peak region, the dimuon mass window is adapted to exclude these backgrounds and the particle identification requirements are loosened to increase the signal yield. The conditions of triggered events are required to be met either from the signal candidate itself or the remainder of the event, which facilitates the modelling of the acceptance. Furthermore the data are analysed in only two bins of the final selection s-BDT, chosen to maximise the sensitivity to the effective lifetime.
The mass distributions in each s-BDT region are fitted independently to extract background-subtracted decay time distributions with the \sPlot~ technique~\cite{sPLOT}. A simultaneous fit to the two background-subtracted decay-time distributions as shown in Fig.~\ref{fig:lifetime_fit_LHCb} is employed to extract the effective lifetime.
In order to extract an unbiased lifetime measurement, the acceptance effects of the reconstruction selection requirements have to be modelled.
The decay time acceptance is modelled by fitting parametric functions to the efficiency distribution in simulation, where the simulation has been weighted to improve data-simulation differences.
The procedure is validated by measuring the lifetimes of \BsKK and \BdKpi in data, finding good agreement with the world average values~\cite{PDG2020}.
The uncertainty of the measurement of the \BsKK lifetime is taken as systematic uncertainty. Further systematic effects like the sample contamination from \mbox{\Bdmumu} and \Bhh decays, acceptance modelling, uncertainties in the background decay time distributions and \Bs-\Bsbar-production asymmetries are investigated and are found to have only sub-leading to negligible effects.
The measured effective lifetime is found to be
\begin{align}
\tau_{\mumu}=2.07\pm0.29\pm0.03\,\text{ps},
\end{align}
where the first uncertainty is statistical and the second systematic.
This value is outside the lifetime interval defined by the \Bs light ($\ADeltaGamma=-1$) and heavy ($\ADeltaGamma=1$) mass eigenstates, but is consistent with these values at the level of 2.2 and 1.5 standard deviations, respectively.
\section{Combination of the measurements by the LHC experiments}
\label{sec:combination}

The latest combination of the measurements from LHC experiments is presented in Ref.~\cite{ATLAS-CMS-LHCb-Combination-2020}. It includes
the results from ATLAS~\cite{Aaboud:2018mst}, CMS~\cite{Sirunyan:2019xdu}, and LHCb~\cite{LHCb-PAPER-2017-001}. The combination is performed based on the two-dimensional profile likelihoods obtained by each experiment from the fits to the dimuon events. 
Such a method allows to properly take into account the correlations between $\BRof{\Bsmumu}$ and $\BRof{\Bdmumu}$, and the upper limit of $\BRof{\Bdmumu}$ can be evaluated using the same inputs. Note the previous combination paper~\cite{CMS:2014xfa} based on CMS and LHCb data collected during LHC Run-1 is based on an unbinned maximum likelihood fit simultaneously to the events from both experiments.

ATLAS results, described in Section \ref{sec:atlas_ana}, are extracted from the data samples of 25~fb$^{-1}$ collected at the center-of-mass energies $\sqrt{s} = $ 7 and 8~TeV, and 26.3~fb$^{-1}$ at $\sqrt{s} = $ 13~TeV. The resulting branching fractions were obtained:
\begin{align}
\BRof{\Bsmumu}&=\left[  2.8^{+0.8}_{-0.7} \right]\times 10^{-9},\\
\BRof{\Bdmumu}&=\left[  -1.9\pm1.6 \right]\times 10^{-10},
\end{align}
where the systematic uncertainties are included in the evaluation. The corresponding significance for $\Bsmumu$ signal is $4.6 \sigma$, 
while the upper limit for $\Bdmumu$ branching fraction is $<2.1\times 10^{-10}$ at 95\% CL.

CMS analysis is based on the data samples of 5 fb$^{-1}$, 20~fb$^{-1}$, and 36~fb$^{-1}$, collected at the center-of-mass energies of $\sqrt{s} = $ 7, 8 and 13~TeV, respectively~\cite{Sirunyan:2019xdu}.  The resulting branching fractions and effective lifetime for $\Bsmumu$ are 
\begin{align}
\BRof{\Bsmumu}&=\left[  2.9^{+0.7}_{-0.6}\,{\rm (exp)}\pm0.2\,{\rm (frag)} \right]\times 10^{-9},\\
\BRof{\Bdmumu}&=\left[  0.8^{+1.4}_{-1.3} \right]\times 10^{-10},\\
\tau_{\Bsmumu} &=1.70^{+0.60}_{-0.43}\,{\rm (stat)} \pm0.09\,{\rm (syst)}~{\rm ps}.
\end{align}
The signals for $\Bsmumu$ and $\Bdmumu$ yield a significance of $5.6 \sigma$ and $1.0 \sigma$, respectively. 
The upper limit for $\Bdmumu$ branching fraction is evaluated to be $<3.6\times 10^{-10}$ at 95\% CL. The first uncertainty of $\BRof{\Bsmumu}$ 
combined statistical and systematic uncertainties from the analysis, while the second uncertainty is from the uncertainty in the fragmentation ratio $f_d/f_s$.

LHCb studies are performed on the data samples of 1 fb$^{-1}$, 2~fb$^{-1}$, and 1.4~fb$^{-1}$, collected at the center-of-mass energies $\sqrt{s} = $ 7, 8 and 13~TeV, respectively~\cite{LHCb-PAPER-2017-001}. The analysis yields the following results:
\begin{align}
\BRof{\Bsmumu}&=\left[  3.0\pm0.6\,{\rm (stat)}^{+0.3}_{-0.2}\,{\rm (syst)} \right]\times 10^{-9},\\
\BRof{\Bdmumu}&=\left[  1.5^{+1.2}_{-1.0}\,{\rm (stat)}^{+0.2}_{-0.1}\,{\rm (syst)} \right]\times 10^{-10},\\
\tau_{\Bsmumu} &=2.04 \pm 0.44\,{\rm (stat)} \pm0.05\,{\rm (syst)}~{\rm ps}, 
\end{align}
with signal significances of $7.8 \sigma$ and $1.6 \sigma$ for $\Bsmumu$ and $\Bdmumu$ decays, respectively.
An upper limit $\BRof{\Bdmumu} < 3.4\times 10^{-10}$ at 95\% CL is obtained.

For the combination of decay branching fractions, profiled likelihoods are computed in the two-dimensional grid of $\BRof{\Bsmumu}$ and $\BRof{\Bdmumu}$ plane from each experiment and the SM $\Bsmumu$ lifetime is assumed. As the current measurements are dominating by statistical uncertainties, the systematic uncertainties are treated independently for the three measurements, except for the common nuisance parameter, $f_d/f_s$ ratio. The $f_d/f_s$ uncertainty is profiled separately in each likelihood and retained only in the LHCb experiment. To test the impact of this correlation, the  $\BRof{\Bsdmumu}$ are evaluated with and without the $f_d/f_s$ uncertainty in ATLAS and CMS liikelihoods. The impact is found to be negligible. Additionally, the dependence of $f_d/f_s$ on the transverse momentum is checked and is found to be consistent within the assigned uncertainties.

\begin{figure}[H]
\includegraphics[width=0.49\textwidth]{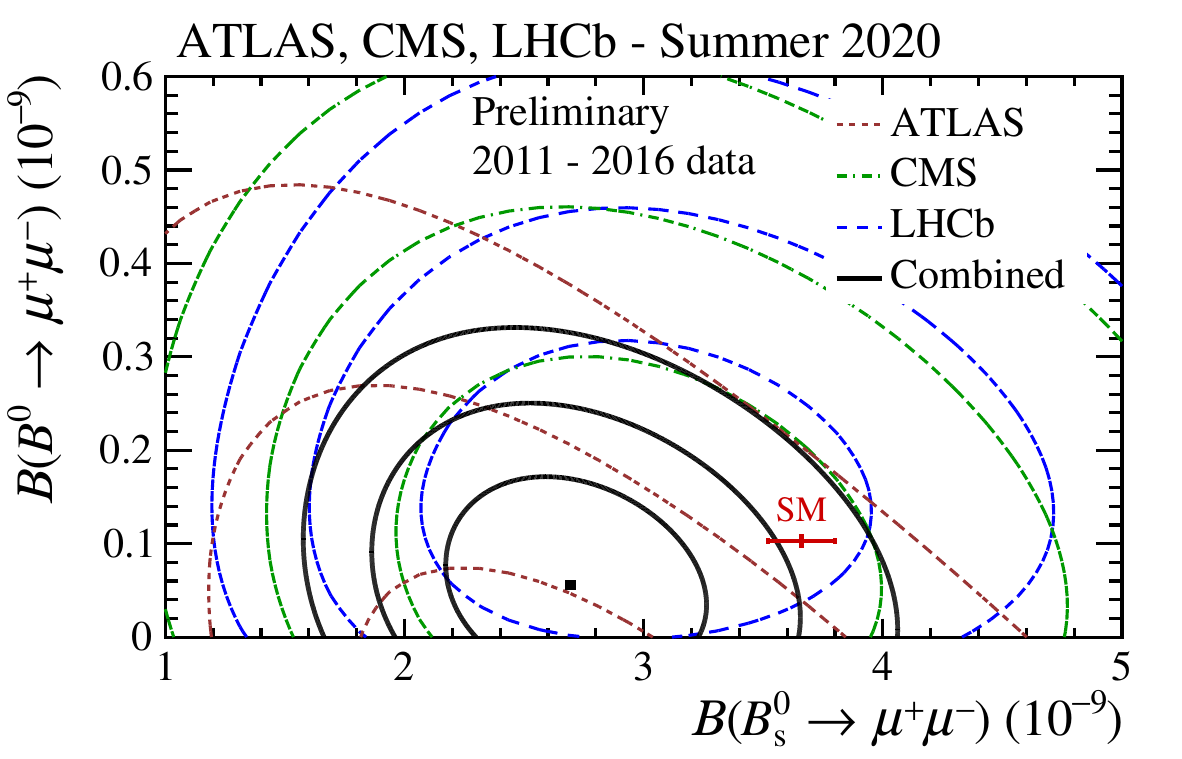}
\includegraphics[width=0.49\textwidth]{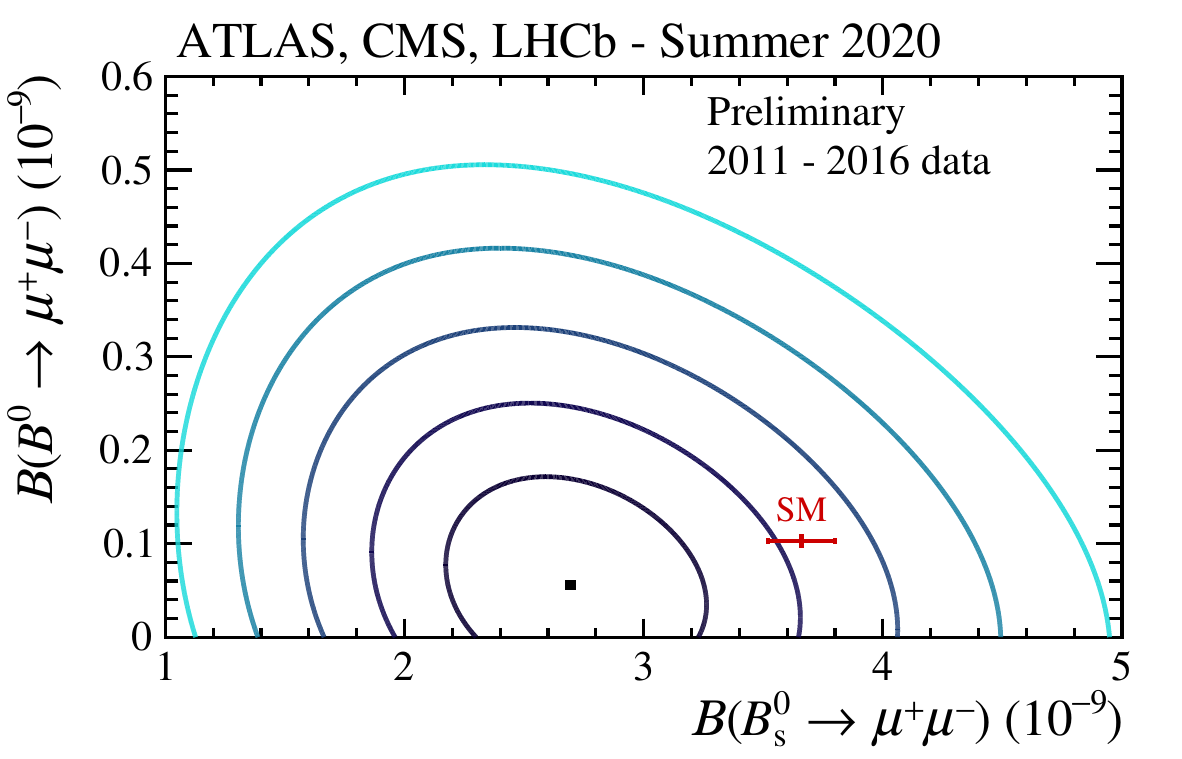}
\caption{ Left plot: the two-dimensional likelihood contours for the $\Bsdmumu$ decays from ATLAS (red dashed line), CMS (green dot-dashed line), and LHCb (blue long-dashed line) experiments, together with contours for their combination (continuous line). The likelihood contours are corresponding to the values of $-2\Delta\ln \mathcal{L}$ = 2.3, 6.2, and 11.8, respectively. Right plot: the likelihood contours for the combination of the three results, corresponding to the values of $-2\Delta\ln \mathcal{L}$ = 2.3, 6.2, 11.8, 19.3, and 30.2, or 1 to 5 $\sigma$ levels in a bidimensional Gaussian approximation.
Figures from Ref.~\cite{ATLAS-CMS-LHCb-Combination-2020}.
\label{fig:combine_2dcontours}}
\end{figure}

The profiled likelihood for each experiment is then modeled with a two-dimensional variable-width Gaussian, which describes asymmetric likelihoods (and asymmetric uncertainties) and also the correlation between the two branching fractions. This analytical function is found to be consistent with the original likelihood for each experiment. The log-likelihoods from the three measurements are summed across the $\BRof{\Bsmumu}$ - $\BRof{\Bdmumu}$ grid points and then fitted using the variable-width Gaussians. By maximizing the modeled likelihood function the combined branching fractions and the associated uncertainties are derived:
 \begin{align}
\BRof{\Bsmumu}&=\left[  2.69^{+0.37}_{-0.35} \right]\times 10^{-9},\\
\BRof{\Bdmumu}&=\left[  0.6 \pm 0.7 \right]\times 10^{-10}.
\end{align}
 The upper limit on $\BRof{\Bdmumu}$ is evaluated as $<1.6 (1.9)\times 10^{-10}$ at 90\% (95\%) CL, which is calculated under the positive $\BRof{\Bdmumu}$ hypothesis by renormalising the likelihood in the interested region. The combined $\BRof{\Bsmumu}$ branching fraction is found to be lower than any single result, which is due to the strong anti-correlation between two branching fractions. 
 The individual profiled likelihood (left) and the combined likelihood in the $\BRof{\Bsmumu}$ - $\BRof{\Bdmumu}$ plane (right) are shown in Fig.~\ref{fig:combine_2dcontours}.
 
The compatibility with the SM predictions is estimated to be 2.4$\sigma$ for $\BRof{\Bsmumu}$, 0.64$\sigma$ for $\BRof{\Bdmumu}$, and 2.1$\sigma$ if 
computed in the $\BRof{\Bsmumu}$ - $\BRof{\Bdmumu}$ plane. These values are calculated assuming Wilks’ theorem and with theoretical uncertainties included. 
In addition to the individual branching fractions, a combined estimation on the ratio of branching fractions $\mathcal{R}$ (see Eq. \ref{eq:RB} is also derived:
 \begin{align}
 \mathcal{R} &= {\BRof{\Bdmumu} \over \BRof{\Bsmumu}} = 0.021^{+0.030}_{-0.025},
\end{align}
where the corresponding upper limit is evaluated to be $\mathcal{R}<0.052$ ($0.060$) at 90\% (95\%) CL.
 
\unskip

The $\Bsmumu$ effective lifetime is measured in the last analysis iteration by all three experiments, as reported in Sections \ref{sec:atlas_ana}--\ref{sec:LHCb_measurements}. However, at the time when the combination was perforemed, only CMS and LHCb Collaborations had a measurement of this quantity. Therefore, a combination has been carried out based only on their results, exploiting a similar method as for the $\BRof{\Bsdmumu}$ combination. The LHCb analysis is carried out with a bin-likelihood fit to the background-subtracted decay time distribution, while the CMS measurement is 
carried out with a two-dimensional likelihood fit to the decay time and dimuon invariant mass distributions. 
As the analyses are fully dominated by the statistical uncertainties, the combination is performed by describing CMS and LHCb likelihoods (as a function of effective lifetime $\tau_{\Bsmumu}$) with variable-width Gaussians, and then, to determine their combined value, the two measurements are assumed to be uncorrelated. 
The resulting $\tau_{\Bsmumu}$ value and the corresponding uncertainty are:
 \begin{align}
\tau_{\Bsmumu} &=1.91^{+0.37}_{-0.35}~{\rm ps}.
\end{align}

Both CMS and LHCb Collaborations have recently released updated analyses, as discussed in Sections \ref{sec:cms_BF}-\ref{sec:LHCb_measurements}. Another iteration of the combination is foreseen in the near future, incorporating the results from all three experiments based on the full Run 2 LHC campaign data.
\section{Conclusion and prospects}
\label{sec:prospects}
In recent years the ATLAS, CMS and LHCb Collaborations made a push towards precision measurements of the \Bsmumu branching fraction, which resulted in measurements that reach a precision of down to $10\,\%$ relative uncertainty. These measurements are the most precise to date. At the same time all three Collaborations have begun measuring the effective lifetime of the decay to understand the \CP structure of the decay. Contrary to initial evidence in the first combination of CMS and LHCb measurements~\cite{CMS:2014xfa}, the \Bdmumu decay has not been confirmed yet. All results are in good agreement with the SM, strongly constraining potential NP scenarios.
To achieve even higher sensitivities, a community effort is ongoing to combine the results of all three experiments. The results of the previous combination have been presented in this review, but have been superseded by the legacy measurements by the CMS and LHCb Collaborations. Once the measurement with the full Run 2 data of the ATLAS Collaboration is published as well, this combination will be repeated to have the most precise picture possible with the harvest of Run 2 data.

After the LHC Run 2, in 2021 the experiments began to take data again with increased instantaneous luminosity until the end of 2025.
After that, the High-Luminosity LHC phase will begin, which will have increased pile-up conditions for all experiments and a massively increased total luminosity.
The ATLAS and CMS experiments will strongly upgrade their detectors to cope with the increased pile-up conditions. However, they also target a significant dimuon mass resolution improvement by $20\,\% - 30\,\%$ (ATLAS) and $40\,\%-50\,\%$ (CMS), respectively.
The LHCb experiment is planning to follow and go through a major upgrade in 2031 to begin taking data with the LHC Run 5.
By the end of the LHC lifetime, ATLAS and CMS aim to have collected $3000\,\invfb$, while LHCb is estimating $300\,\invfb$.
Under these conditions and assuming the central values as predicted by the Standard Model, the ATLAS, CMS and LHCb collaborations made extrapolations to the expected sensitivity of future measurements~\cite{ATLAS:2022hsp, LHCb:2018hne}.
For the ATLAS experiment, the sensitivity strongly depends on the trigger conditions for dimuon events with the upgraded detector.
In the most conservative scenario the expected statistical-only (statistical and systematic) uncertainties reach $19\,\%$ ($23\,\%$) relative to the central value for the \Bsmumu branching fraction and $134\,\%$ ($135\,\%$) for the \Bdmumu branching fraction, while in the most optimistic scenario they reach $5\,\%$ ($13$) for \Bsmumu and $25\,\%$ ($26\,\%$) for \Bdmumu.
The dominant systematic uncertainties in these projections arise from external inputs like the uncertainty on the fragmentation fraction ratio $f_s/f_d$ and the branching fraction uncertainty of the normalisation channel.

The CMS collaboration expects to reach uncertainties of $7\,\%$ on the branching fraction of \Bsmumu and $16\,\%$ on the branching fraction of \Bdmumu.
The expected uncertainty on the effective \Bsmumu lifetime is $0.05\,$ps.
This precision will allow stringent constraints on the parameter $A^{\mumu}_{\Delta\Gamma}$ and in particular break the degeneracy between possible scalar and pseudoscalar contributions beyond the SM to this decay.

The LHCb collaboration expects to reach a statistical uncertainty on the \Bsmumu branching fraction of $1.8\%$, however, the analysis will be systematically limited by the external inputs of the fragmentation fraction ratios and the normalisation branching fractions, which are estimated to become $4\,\%$ by then. On the contrary, the ratio $\mathcal{B}(\Bdmumu)/\mathcal{B}(\Bsmumu)$ is not expected to become systematically limited and is expected to reach a relative precision of $10\,\%$.
The measurement of the effective \Bsmumu lifetime is expected to reach a precision of $0.033\,$ps.
Both the CMS and LHCb Collaborations expect to establish the \Bdmumu decay signal at more than $5\sigma$ level. 

The expected large yield of \Bsmumu decays will also allow to access the \CP parameter $S_{\mumu}$, which describes the time-dependent \CP-violation in the decay~\cite{Buras:2013uqa}. Adding this parameter will complete the base of \CP observables and provide complementary constraints to physics beyond the SM that are not constrained by the other observables.
A nonzero value of this parameter will be an immediate sign for a \CP-violating phase beyond the SM.
This parameter can only be determined by measuring the decay-time distribution of \Bs and \Bsb decays separately and thus requires the tagging of the \Bs flavour.
Assuming a similar performance of the flavour tagging as in Run 2, the LHCb collaboration expects to reach a precision of $0.2$ of this parameter. Provided a sufficient flavour tagging performance can be achieved, this analysis could potentially be performed by the CMS and ATLAS experiments.

To achieve the projected sensitivities discussed in this section and possibly surpass them, it will be important to maintain the basic assumptions. For the ATLAS and CMS experiments it will be crucial to design trigger strategies that allow to keep the muon transverse momentum thresholds as low as possible in the high pile-up environment. Furthermore the level of backgrounds from random combinations must be maintained or decreased, which might be achieved through the tracking detectors, the fast timing information in the reconstruction and the improvement of current selection algorithms based on Machine Learning tools.
Fast timing information to disentangle $pp$-collision points will also facilitate the analysis of LHCb data and enable the flavour tagging of the \Bs mesons.
Further improvements over the projected sensitivities in this section - especially on \Bdmumu measurements - might be achieved by improvements on the muon identification and the momentum resolution, which will have significant impact on the dimuon mass resolution.

\acknowledgments{The authors thank the ATLAS, CMS and LHCb physics working groups for cross reading the draft. K.-F. Chen is supported by the grant 112-2112-M-002-026 of National Science and Technology Council, Taiwan.}





\appendixtitles{no} 



\begin{adjustwidth}{-\extralength}{0cm}

\reftitle{References}


\bibliography{main.bib}

\begin{thebibliography}{999}

\bibitem[Bobeth et~al.(2014{\natexlab{a}})Bobeth, Gorbahn, Hermann, Misiak,
  Stamou, et~al.]{Bobeth:2013uxa}
Bobeth, C.; Gorbahn, M.; Hermann, T.; Misiak, M.; Stamou, E.;  et~al.
\newblock {$B_{s,d} \to \ell^+ \ell^-$ in the Standard Model with Reduced
  Theoretical Uncertainty}.
\newblock {\em Phys.Rev.Lett.} {\bf 2014}, {\em 112},~101801,
  \href{https://arxiv.org/abs/1311.0903}{{\normalfont
  [arXiv:hep-ph/1311.0903]}}.
\newblock {\url{https://doi.org/10.1103/PhysRevLett.112.101801}}.

\bibitem[Bobeth et~al.(2014{\natexlab{b}})Bobeth, Gorbahn, and
  Stamou]{Bobeth:2013tba}
Bobeth, C.; Gorbahn, M.; Stamou, E.
\newblock {Electroweak Corrections to $B_{s,d} \to \ell^+ \ell^-$}.
\newblock {\em Phys. Rev. D} {\bf 2014}, {\em 89},~034023,
  \href{https://arxiv.org/abs/1311.1348}{{\normalfont
  [arXiv:hep-ph/1311.1348]}}.
\newblock {\url{https://doi.org/10.1103/PhysRevD.89.034023}}.

\bibitem[Hermann et~al.(2013)Hermann, Misiak, and Steinhauser]{Hermann:2013kca}
Hermann, T.; Misiak, M.; Steinhauser, M.
\newblock {Three-loop QCD corrections to $B_s \to \mu^+ \mu^-$}.
\newblock {\em JHEP} {\bf 2013}, {\em 12},~097,
  \href{https://arxiv.org/abs/1311.1347}{{\normalfont
  [arXiv:hep-ph/1311.1347]}}.
\newblock {\url{https://doi.org/10.1007/JHEP12(2013)097}}.

\bibitem[Beneke et~al.(2018)Beneke, Bobeth, and Szafron]{Beneke:2017vpq}
Beneke, M.; Bobeth, C.; Szafron, R.
\newblock {Enhanced electromagnetic correction to the rare B-meson decay
  $B_{s,d} \to \mu^+ \mu^-$}.
\newblock {\em Phys. Rev. Lett.} {\bf 2018}, {\em 120},~011801,
  \href{https://arxiv.org/abs/1708.09152}{{\normalfont
  [arXiv:hep-ph/1708.09152]}}.
\newblock {\url{https://doi.org/10.1103/PhysRevLett.120.011801}}.

\bibitem[Beneke et~al.(2019)Beneke, Bobeth, and Szafron]{Beneke:2019slt}
Beneke, M.; Bobeth, C.; Szafron, R.
\newblock {Power-enhanced leading-logarithmic QED corrections to $B_q \to
  \mu^+\mu^-$}.
\newblock {\em JHEP} {\bf 2019}, {\em 10},~232,
  \href{https://arxiv.org/abs/1908.07011}{{\normalfont
  [arXiv:hep-ph/1908.07011]}}.
\newblock {\url{https://doi.org/10.1007/JHEP10(2019)232}}.

\bibitem[Altmannshofer et~al.(2012)Altmannshofer, Paradisi, and
  Straub]{Altmannshofer:2011gn}
Altmannshofer, W.; Paradisi, P.; Straub, D.M.
\newblock {Model-Independent Constraints on New Physics in $b \to s$
  Transitions}.
\newblock {\em JHEP} {\bf 2012}, {\em 04},~008,
  \href{https://arxiv.org/abs/1111.1257}{{\normalfont
  [arXiv:hep-ph/1111.1257]}}.
\newblock {\url{https://doi.org/10.1007/JHEP04(2012)008}}.

\bibitem[Beaujean et~al.(2012)Beaujean, Bobeth, van Dyk, and
  Wacker]{Beaujean:2012uj}
Beaujean, F.; Bobeth, C.; van Dyk, D.; Wacker, C.
\newblock {Bayesian Fit of Exclusive $b \to s \bar\ell\ell$ Decays: The
  Standard Model Operator Basis}.
\newblock {\em JHEP} {\bf 2012}, {\em 08},~030,
  \href{https://arxiv.org/abs/1205.1838}{{\normalfont
  [arXiv:hep-ph/1205.1838]}}.
\newblock {\url{https://doi.org/10.1007/JHEP08(2012)030}}.

\bibitem[Aoki et~al.(2019)]{Aoki:2019cca}
Aoki, S.;  et~al.
\newblock {FLAG Review 2019},
  \href{https://arxiv.org/abs/1902.08191}{{\normalfont
  [arXiv:hep-lat/1902.08191]}}.

\bibitem[Bazavov et~al.(2018)]{Bazavov:2017lyh}
Bazavov, A.;  et~al.
\newblock {B- and D-meson leptonic decay constants from four-flavor lattice
  QCD}.
\newblock {\em Phys. Rev. D} {\bf 2018}, {\em 98},~074512,
  \href{https://arxiv.org/abs/1712.09262}{{\normalfont
  [arXiv:hep-lat/1712.09262]}}.
\newblock {\url{https://doi.org/10.1103/PhysRevD.98.074512}}.

\bibitem[Bussone et~al.(2016)]{Bussone:2016iua}
Bussone, A.;  et~al.
\newblock {Mass of the b quark and B meson decay constants from N$_f$=2+1+1
  twisted-mass lattice QCD}.
\newblock {\em Phys. Rev. D} {\bf 2016}, {\em 93},~114505,
  \href{https://arxiv.org/abs/1603.04306}{{\normalfont
  [arXiv:hep-lat/1603.04306]}}.
\newblock {\url{https://doi.org/10.1103/PhysRevD.93.114505}}.

\bibitem[Dowdall et~al.(2013)Dowdall, Davies, Horgan, Monahan, and
  Shigemitsu]{Dowdall:2013tga}
Dowdall, R.J.; Davies, C.T.H.; Horgan, R.R.; Monahan, C.J.; Shigemitsu, J.
\newblock {B-meson decay constants from improved lattice nonrelativistic QCD
  with physical u, d, s, and c quarks}.
\newblock {\em Phys. Rev. Lett.} {\bf 2013}, {\em 110},~222003,
  \href{https://arxiv.org/abs/1302.2644}{{\normalfont
  [arXiv:hep-lat/1302.2644]}}.
\newblock {\url{https://doi.org/10.1103/PhysRevLett.110.222003}}.

\bibitem[Hughes et~al.(2018)Hughes, Davies, and Monahan]{Hughes:2017spc}
Hughes, C.; Davies, C.T.H.; Monahan, C.J.
\newblock {New methods for B meson decay constants and form factors from
  lattice NRQCD}.
\newblock {\em Phys. Rev. D} {\bf 2018}, {\em 97},~054509,
  \href{https://arxiv.org/abs/1711.09981}{{\normalfont
  [arXiv:hep-lat/1711.09981]}}.
\newblock {\url{https://doi.org/10.1103/PhysRevD.97.054509}}.

\bibitem[De~Bruyn et~al.(2012)De~Bruyn, Fleischer, Knegjens, Koppenburg, Merk,
  et~al.]{DeBruyn:2012wk}
De~Bruyn, K.; Fleischer, R.; Knegjens, R.; Koppenburg, P.; Merk, M.;  et~al.
\newblock {Probing new physics via the $B^0_s\to \mumu$ effective lifetime}.
\newblock {\em Phys.Rev.Lett.} {\bf 2012}, {\em 109},~041801,
  \href{https://arxiv.org/abs/1204.1737}{{\normalfont
  [arXiv:hep-ph/1204.1737]}}.
\newblock {\url{https://doi.org/10.1103/PhysRevLett.109.041801}}.

\bibitem[Bruyn et~al.(2012)Bruyn, Fleischer, Knegjens, Koppenburg, Merk,
  et~al.]{DeBruyn:2012wj}
Bruyn, K.D.; Fleischer, R.; Knegjens, R.; Koppenburg, P.; Merk, M.;  et~al.
\newblock {Branching Ratio Measurements of \Bs\ Decays}.
\newblock {\em Phys.Rev.} {\bf 2012}, {\em D86},~014027,
  \href{https://arxiv.org/abs/1204.1735}{{\normalfont
  [arXiv:hep-ph/1204.1735]}}.
\newblock {\url{https://doi.org/10.1103/PhysRevD.86.014027}}.

\bibitem[Buras(2003)]{Buras:2003td}
Buras, A.J.
\newblock {Relations between $\Delta$ M($s$, $d)$ and B($s$, $d) \to \mu
  \bar{\mu}$ in models with minimal flavor violation}.
\newblock {\em Phys. Lett. B} {\bf 2003}, {\em 566},~115--119,
  \href{https://arxiv.org/abs/hep-ph/0303060}{{\normalfont [hep-ph/0303060]}}.
\newblock {\url{https://doi.org/10.1016/S0370-2693(03)00561-6}}.

\bibitem[King et~al.(2019)King, Lenz, and Rauh]{King:2019lal}
King, D.; Lenz, A.; Rauh, T.
\newblock {B$_{s}$ mixing observables and |V$_{td}$/V$_{ts}$| from sum rules}.
\newblock {\em JHEP} {\bf 2019}, {\em 05},~034,
  \href{https://arxiv.org/abs/1904.00940}{{\normalfont
  [arXiv:hep-ph/1904.00940]}}.
\newblock {\url{https://doi.org/10.1007/JHEP05(2019)034}}.

\bibitem[Buras(2023)]{Buras:2022}
Buras, A.J.
\newblock {Standard Model Predictions for Rare K and B Decays without New
  Physics Infection}.
\newblock {\em EPJC} {\bf 2023}, {\em 83},~66,
  \href{https://arxiv.org/abs/2209.03968}{{\normalfont
  [arXiv:hep-ph/2209.03968]}}.
\newblock {\url{https://doi.org/10.1140/epjc/s10052-023-11222-6}}.

\bibitem[Amhis et~al.(2023)]{HFLAV:2022pwe}
Amhis, Y.;  et~al.
\newblock {Averages of $b$-hadron, $c$-hadron, and $\tau$-lepton properties as
  of 2021}.
\newblock {\em Phys. Rev. D} {\bf 2023}, {\em 107},~052008,
  \href{https://arxiv.org/abs/2206.07501}{{\normalfont
  [arXiv:hep-ex/2206.07501]}}.
\newblock {\url{https://doi.org/10.1103/PhysRevD.107.052008}}.

\bibitem[{ATLAS Collaboration}(2019)]{Aaboud:2018mst}
{ATLAS Collaboration}.
\newblock {Study of the rare decays of $B^0_s$ and $B^0$ mesons into muon pairs
  using data collected during 2015 and 2016 with the ATLAS detector}.
\newblock {\em JHEP} {\bf 2019}, {\em 04},~098,
  \href{https://arxiv.org/abs/1812.03017}{{\normalfont
  [arXiv:hep-ex/1812.03017]}}.
\newblock {\url{https://doi.org/10.1007/JHEP04(2019)098}}.

\bibitem[{ATLAS Collaboration}(2016)]{Aaboud:2016ire}
{ATLAS Collaboration}.
\newblock {Study of the rare decays of $B^0_s$ and $B^0$ into muon pairs from
  data collected during the LHC Run 1 with the ATLAS detector}.
\newblock {\em Eur. Phys. J. C} {\bf 2016}, {\em 76},~513,
  \href{https://arxiv.org/abs/1604.04263}{{\normalfont
  [arXiv:hep-ex/1604.04263]}}.
\newblock {\url{https://doi.org/10.1140/epjc/s10052-016-4338-8}}.

\bibitem[Hoecker et~al.(2007)Hoecker, Speckmayer, Stelzer, Therhaag, von
  Toerne, and Voss]{Hocker:2007ht}
Hoecker, A.; Speckmayer, P.; Stelzer, J.; Therhaag, J.; von Toerne, E.; Voss,
  H.
\newblock {TMVA: Toolkit for Multivariate Data Analysis}.
\newblock {\em PoS} {\bf 2007}, {\em ACAT},~040,
  \href{https://arxiv.org/abs/physics/0703039}{{\normalfont
  [physics/0703039]}}.

\bibitem[Tanabashi et~al.(2018)]{PDG2018}
Tanabashi, M.;  et~al.
\newblock {Review of Particle Physics}.
\newblock {\em Phys. Rev. D} {\bf 2018}, {\em 98},~030001.
\newblock {\url{https://doi.org/10.1103/PhysRevD.98.030001}}.

\bibitem[Amhis et~al.(2017)]{HFLAV2016}
Amhis, Y.;  et~al.
\newblock {Averages of $b$-hadron, $c$-hadron, and $\tau$-lepton properties as
  of summer 2016}.
\newblock {\em Eur. Phys. J. C} {\bf 2017}, {\em 77},~895,
  \href{https://arxiv.org/abs/1612.07233}{{\normalfont
  [arXiv:hep-ex/1612.07233]}}.
\newblock {\url{https://doi.org/10.1140/epjc/s10052-017-5058-4}}.

\bibitem[Neyman()]{neyman}
Neyman, J.
\newblock Outline of a Theory of Statistical Estimation Based on the Classical
  Theory of Probability.
\newblock \href{http://rsta.royalsocietypublishing.org/content/236/767/333}
  {Phil. Trans. R. Soc. London A, \textbf{236} (1937) 333-380}.

\bibitem[{ATLAS Collaboration}(2023)]{lifetime_ATLAS}
{ATLAS Collaboration}.
\newblock {Measurement of the $B_s^0 \to \mu\mu$ Effective Lifetime with the
  ATLAS Detector}.
\newblock {\em JHEP} {\bf 2023}, {\em 09},~199,
  \href{https://arxiv.org/abs/2308.01171}{{\normalfont
  [arXiv:hep-ex/2308.01171]}}.
\newblock {\url{https://doi.org/10.1007/JHEP09(2023)199}}.

\bibitem[Workman and Others(2022)]{PDG_2022}
Workman, R.L.; Others.
\newblock {Review of Particle Physics}.
\newblock {\em Progr. Theor. Exp. Phys.} {\bf 2022}, {\em 2022},~083C01 and
  2023 update.
\newblock {\url{https://doi.org/10.1093/ptep/ptac097}}.

\bibitem[Pivk and Le~Diberder(2005)]{sPLOT}
Pivk, M.; Le~Diberder, F.
\newblock {sPlot: A statistical tool to unfold data distributions}.
\newblock {\em Nucl. Instrum. Meth. A} {\bf 2005}, {\em 555},~356--369,
  \href{https://arxiv.org/abs/physics/0402083v3}{{\normalfont
  [physics/0402083v3]}}.
\newblock {\url{https://doi.org/https://doi.org/10.1016/j.nima.2005.08.106}}.

\bibitem[{CMS Collaboration}(2022)]{CMS-BPH-21-006}
{CMS Collaboration}.
\newblock {Measurement of the \(B^0_{s} \to \mu^+\mu^-\) decay properties and
  search for the \(B^0 \to \mu^+\mu^-\) decay in proton--proton collisions at
  \(\sqrt{s} = 13\,\text{TeV}\)}.
\newblock {\em Phys. Lett. B} {\bf 2022}, {\em 842},~137955,
  \href{https://arxiv.org/abs/2212.10311}{{\normalfont
  [arXiv:hep-ex/2212.10311]}}.
\newblock {\url{https://doi.org/10.1016/j.physletb.2023.137955}}.

\bibitem[{CMS Collaboration}(2020)]{CMS-BPH-16-004}
{CMS Collaboration}.
\newblock {Measurement of properties of \(B^0_{s} \to \mu^+\mu^-\) decays and
  search for \(B^0 \to \mu^+\mu^-\) with the CMS experiment}.
\newblock {\em JHEP} {\bf 2020}, {\em 04},~188,
  \href{https://arxiv.org/abs/1910.12127}{{\normalfont
  [arXiv:hep-ex/1910.12127]}}.
\newblock {\url{https://doi.org/10.1007/JHEP04(2020)188}}.

\bibitem[Chen and Guestrin(2016)]{Chen:2016btl}
Chen, T.; Guestrin, C.
\newblock {XGBoost: A Scalable Tree Boosting System} {\bf 2016}.
\newblock  \href{https://arxiv.org/abs/1603.02754}{{\normalfont
  [arXiv:cs.LG/1603.02754]}}.
\newblock {\url{https://doi.org/10.1145/2939672.2939785}}.

\bibitem[Pivk and Le~Diberder(2005)]{Pivk:2004ty}
Pivk, M.; Le~Diberder, F.R.
\newblock {sPlot: a statistical tool to unfold data distributions}.
\newblock {\em Nucl.Instrum.Meth.} {\bf 2005}, {\em A555},~356--369,
  \href{https://arxiv.org/abs/physics/0402083}{{\normalfont
  [arXiv:physics.data-an/physics/0402083]}}.
\newblock {\url{https://doi.org/10.1016/j.nima.2005.08.106}}.

\bibitem[Aaij et~al.(2021)]{LHCb:2021qbv}
Aaij, R.;  et~al.
\newblock {Precise measurement of the~$f_s/f_d$ ratio of fragmentation
  fractions and of $B^0_s$ decay branching fractions}.
\newblock {\em Phys. Rev. D} {\bf 2021}, {\em 104},~032005,
  \href{https://arxiv.org/abs/2103.06810}{{\normalfont
  [arXiv:hep-ex/2103.06810]}}.
\newblock {\url{https://doi.org/10.1103/PhysRevD.104.032005}}.

\bibitem[Read(2002)]{Read:2002hq}
Read, A.L.
\newblock {Presentation of search results: The CL(s) technique}.
\newblock {\em J.Phys.} {\bf 2002}, {\em G28},~2693--2704.
\newblock {\url{https://doi.org/10.1088/0954-3899/28/10/313}}.

\bibitem[Aaij et~al.(2022{\natexlab{a}})]{LHCb:2021awg}
Aaij, R.;  et~al.
\newblock {Measurement of the $B^0_s\to\mu^+\mu^-$ decay properties and search
  for the $B^0\to\mu^+\mu^-$ and $B^0_s\to\mu^+\mu^-\gamma$ decays}.
\newblock {\em Phys. Rev. D} {\bf 2022}, {\em 105},~012010,
  \href{https://arxiv.org/abs/2108.09283}{{\normalfont
  [arXiv:hep-ex/2108.09283]}}.
\newblock {\url{https://doi.org/10.1103/PhysRevD.105.012010}}.

\bibitem[Aaij et~al.(2022{\natexlab{b}})]{LHCb:2021vsc}
Aaij, R.;  et~al.
\newblock {Analysis of Neutral B-Meson Decays into Two Muons}.
\newblock {\em Phys. Rev. Lett.} {\bf 2022}, {\em 128},~041801,
  \href{https://arxiv.org/abs/2108.09284}{{\normalfont
  [arXiv:hep-ex/2108.09284]}}.
\newblock {\url{https://doi.org/10.1103/PhysRevLett.128.041801}}.

\bibitem[Aaij et~al.(2017)]{LHCb-PAPER-2017-001}
Aaij, R.;  et~al.
\newblock {Measurement of the $\Bs\to \mup\mun$ branching fraction and
  effective lifetime and search for $\Bd\to \mup\mun$ decays}.
\newblock {\em Phys. Rev. Lett.} {\bf 2017}, {\em 118},~191801,
  \href{https://arxiv.org/abs/1703.05747}{{\normalfont
  [arXiv:hep-ex/1703.05747]}}.
\newblock {\url{https://doi.org/10.1103/PhysRevLett.118.191801}}.

\bibitem[Archilli et~al.(2013)]{Archilli:2013npa}
Archilli, F.;  et~al.
\newblock {Performance of the muon identification at LHCb}.
\newblock {\em JINST} {\bf 2013}, {\em 8},~P10020,
  \href{https://arxiv.org/abs/1306.0249}{{\normalfont
  [arXiv:physics.ins-det/1306.0249]}}.
\newblock {\url{https://doi.org/10.1088/1748-0221/8/10/P10020}}.

\bibitem[Zyla et~al.(2020)]{PDG2020}
Zyla, P.A.;  et~al.
\newblock {Review of Particle Physics}.
\newblock {\em PTEP} {\bf 2020}, {\em 2020},~083C01.
\newblock {\url{https://doi.org/10.1093/ptep/ptaa104}}.

\bibitem[Skwarnicki(1986)]{Skwarnicki:1986xj}
Skwarnicki, T.
\newblock {A study of the radiative cascade transitions between the
  Upsilon-prime and Upsilon resonances}.
\newblock PhD thesis, Institute of Nuclear Physics, Krakow,  1986.
\newblock
  {\href{http://inspirehep.net/record/230779/files/230779.pdf}{DESY-F31-86-02}}.

\bibitem[ATL(2020)]{ATLAS-CMS-LHCb-Combination-2020}
{Combination of the ATLAS, CMS and LHCb results on the $B^0_{(s)} \to
  \mu^+\mu^-$ decays}.
\newblock Technical report, CERN,  2020.
\newblock ATLAS-CONF-2020-049, CMS-PAS-BPH-20-003, LHCb-CONF-2020-002.

\bibitem[Sirunyan et~al.(2019)]{Sirunyan:2019xdu}
Sirunyan, A.M.;  et~al.
\newblock {Measurement of properties of B$^0_\mathrm{s}\to\mu^+\mu^-$ decays
  and search for B$^0\to\mu^+\mu^-$ with the CMS experiment} {\bf 2019}.
\newblock  \href{https://arxiv.org/abs/1910.12127}{{\normalfont
  [arXiv:hep-ex/1910.12127]}}.

\bibitem[Khachatryan et~al.(2015)]{CMS:2014xfa}
Khachatryan, V.;  et~al.
\newblock {Observation of the rare $B^0_s\to\mu^+\mu^-$ decay from the combined
  analysis of CMS and LHCb data}.
\newblock {\em Nature} {\bf 2015}, {\em 522},~68--72,
  \href{https://arxiv.org/abs/1411.4413}{{\normalfont
  [arXiv:hep-ex/1411.4413]}}.
\newblock {\url{https://doi.org/10.1038/nature14474}}.

\bibitem[ATL(2022)]{ATLAS:2022hsp}
{Snowmass White Paper Contribution: Physics with the Phase-2 ATLAS and CMS
  Detectors} {\bf 2022}.

\bibitem[LHC(2018)]{LHCb:2018hne}
{Physics case for an LHCb Upgrade II}: {Opportunities in flavour physics, and
  beyond, in the HL-LHC era} {\bf 2018}.

\bibitem[Buras et~al.(2013)Buras, Fleischer, Girrbach, and
  Knegjens]{Buras:2013uqa}
Buras, A.J.; Fleischer, R.; Girrbach, J.; Knegjens, R.
\newblock {Probing New Physics with the $B_s \to {\mu}+ {\mu}-$ Time-Dependent
  Rate}.
\newblock {\em JHEP} {\bf 2013}, {\em 07},~077,
  \href{https://arxiv.org/abs/1303.3820}{{\normalfont
  [arXiv:hep-ph/1303.3820]}}.
\newblock {\url{https://doi.org/10.1007/JHEP07(2013)077}}.

\end{thebibliography}

\PublishersNote{}
\end{adjustwidth}
\end{document}